\magnification1200


\vskip 2cm
\centerline
{\bf The massless irreducible representation in E theory and how bosons can appear as  spinors.}
\vskip 1cm
\centerline{Keith Glennon and Peter West}
\centerline{Department of Mathematics}
\centerline{King's College, London WC2R 2LS, UK}
\vskip 2cm
\leftline{\sl Abstract}  

\noindent

\vskip .5cm
We study in detail the irreducible representation of E theory that corresponds to  massless particles. This has little algebra $I_c(E_9)$ and contains 128 physical states that belong to the spinor representation of SO(16). These are the degrees of freedom of maximal supergravity in eleven dimensions. This  smaller number of the degrees of freedom, compared to what might be expected,  is due to an infinite number of duality relations which in turn can be traced to the existence of a subaglebra of $I_c(E_9)$  which forms an ideal and annihilates the representation. We explain how  these features are inherited into the covariant theory. We also comment on the remarkable similarity between how the bosons and fermions arise in E theory. 
\vfill
\eject

\medskip
{{\bf 1. Introduction}}
\medskip
The symmetries of the non-linear realisation of $E_{11} \otimes_s l_1$ with respect to  the Cartan involution invariant subalgebra of $E_{11}$, denoted by  $I_c(E_{11})$, lead to equations of motion at low levels  that   are precisely those of maximal supergravity provided one 
discards the dependence on the coordinates of the spacetime that are beyond those we usually consider [1,2,3,4]. In particular the degrees of freedom they contain are those of the familiar graviton and the three form. While there is no  complete understanding of the role of all the higher level fields  a large class of them are known to  provide different field descriptions of the  graviton and the three form degrees and freedom to which they are related by  invariant duality relations. The higher level fields  fields also  account for the gauging of the maximal supergravity theories. Indeed the non-linear realisation contains all the maximal supergravity theories in their different dimensions including those that are gauged. 
For a review see references [5,6]. 
\par
In a different approach the irreducible representation  of the semi-direct product of $I_c(E_{11})$ with its vector representation $l_1$, denoted $I_c(E_{11})\otimes_s l_1 $ were formulated [7]. This latter algebra is the analogue of the Poincare group which can be written in the form 
$SO(1,D-1)\otimes _s T^D$ where $T^D$ are the translations in $D$ dimensions. The irreducible representations of the Poincare group  were found in 1939 by Wigner [8] and one can use a similar method in E theory. One important difference is that while the construction of the  irreducible representations of the Poincare algebra begin by considering the possible values of the momentum, those in E theory involve the vector, or first fundamental representation of $E_{11}$  and these include all brane charges [2,9,10,11,12]. Thus while the irreducible representations of the Poincare algebra lead to all possible particles those in E theory lead to all possible point particle and branes, that is, all extended objects in E theory. 
\par
The irreducible representation that arises when all the members of the vector representation vanish except for the usual momentum, which was taken to be massless, were studied briefly in reference [7]. It was found that the corresponding little group was $I_c(E_9)$ and it was argued  that the representation only contains a finite number of states which are those of  the graviton and three form in eleven dimensions and so the degrees of freedom of eleven dimensional supergravity. In this paper we will study this irreducible representation in detail. We will show that one can impose an infinite number of duality equations on the representation that are invariant under $I_c(E_9)$ and reduce the number of  independent fields to be 128. These are the fields  $h_{i_1i_2}=h_{(i_1i_2)}$, $ h^i{}_i=0$ and $A_{i_1i_2i_3}= A_{[i_1i_2i_3]}$,  $i_1, i_2,\ldots =2,\ldots , 10$ and they  belong to the 128 dimensional spinor representation of $I_c(E_8)={\rm SO}(16)$. Corresponding to the duality relations we find that there exist an infinite number of generators of $I_c(E_9)$  which annihilate the representation and these  form a subalgebra $I$ that is an ideal. Indeed the Lie algebra ${I_c(E_9)\over I}= {\rm SO}(16)$. 
\par
To understand how it is that the bosonic states, and in particular the graviton,  can belong to the spinor representation of an algebra that involves spacetime symmetries we will decompose the physical states into representations of the subalgebra ${\rm SO(8)}\otimes {\rm SO(8)}$ of ${\rm SO}(16)$. We find that they belong to the representations $128= (8_v, 8_v) \oplus (8_c ,  8_s)$. The graviton belongs to the first representation. 
\par
Given an irreducible representation of the Poincare algebra one can  embed it in a larger representation to find a Lorentz covariant formulation of the representation. The price is that the Lorentz covariant fields obey conditions which are the physical sate conditions and are subject to gauge transformations. We carry out the same procedure for the massless irreducible representation of $I_c(E_{11})\otimes_s l_1 $ and make contact with the non-linear realisation of $E_{11} \otimes_s l_1$ with respect to  the Cartan involution invariant subalgebra of $E_{11}$, denoted by  $I_c(E_{11})$. Indeed the degrees of freedom of the massless  irreducible representation are the ones of the latter theory. This supports the asscertion that these are the only degrees of freedom that arise in the non-linear realisation and it explains the origin and structure of the duality relations that arise in the non-linear realisation.

\medskip
{{\bf 2. The irreducible representation for a  massless particle in E theory}}
\medskip
The irreducible representations in E theory were discussed in reference [7]. To be more precise this paper  studied the irreducible representations of 
the semi-direct product of $I_c(E_{11})$ and it's vector ($l_1$) representation, denoted $I_c(E_{11}) \otimes_s l_1$.  This is a natural extension of the method used to find the  irreducible representations  of the Poincare group  [8]. The similarity comes from the fact that the Poincare group is a semi-direct product of the Lorentz group $SO(1,3)$ and the group of translations $T_4$, denoted $SO(1,3) \otimes_s T_4$
Indeed at the lowest level $I_c(E_{11}) \otimes_s l_1$ is just the Poincare group in eleven dimensions. For the case of the Poincare group one selects a value for the momentum corresponding to if it is  massive or massless and then one computes the little group that preserves this choice. One then takes a representation of this little group and finds a representation of the full Poincare group by boosting,  or  said more technically, by taking an induced representation. In this way one can find all irreducible representations of the Poincare group. 
\par
In E theory the vector ($l_1$)  representation contains all the brane charges and so the first step is to select a preferred value of the vector representation and compute the little algebra  that preserves this value. One then takes an irreducible representation of this little algebra  and boosts it to the full $I_c(E_{11}) \otimes_s l_1$ algebra. At lowest levels the vector representation contains $P_{a}, Z^{a_1a_2}, Z^{a_1\ldots a_5},  \ldots $ with $a_1, a_2,... = 0,1,...,10$ where $P_a$ are just the usual momenta and  $Z^{a_1a_2}$ and $ Z^{a_1\ldots a_5}$ the well known two form and five form charges that first appeared in the supersymmetry algebra. In reference [7] the irreducible representation that arises when one takes the momentum to take a value corresponding to it being massless with the other components of the vector representation  being zero was discussed. We will refer to this irreducible representation as the massless irreducible  representation. Taking some of the higher  charges to be  non-zero one finds the irreducible representations corresponding to branes. 
\par
The purpose of this section is to fully elucidate the properties of the massless irreducible representation. To begin the construction we will take the momentum to have the values $p_0 = - m, p_{10} = m$ with all other components being zero. It will be  more convenient to use  light-cone  notation, for which the components of a vector $V^{a}$ are defined as $V^{\pm} = {1 \over \sqrt{2}} (V^{10} \pm V^0), V^i, 1,...,9$, and the Minkowski metric becomes $\eta_{+-} = 1, \eta_{ij} = \delta_{ij}$ so that $V_{\pm} = {1 \over \sqrt{2}} (V_{10} \pm V_0)$. In light-cone notation the massless irreducible representation begins by taking  the brane charges to be  $p^+= p_- = \sqrt{2} m$, with all other momenta  and other brane charges being set equal to zero.
We next seek the subalgebra ${\cal H}$ of $I_c(E_{11})$ which preserves this  choice of brane charges. The brane charges $l_A$ transform under $I_c(E_{11})$ and so the subalgebra ${\cal H}$ is determined by the requirement that 
$$
\delta l_A = [\Lambda^{\underline{\alpha}} S_{\underline{\alpha}},l_A] = 0
 \eqno(2.1)$$
where the charges $l_A$ take the above values
\par
It is straightforward to show that for a massless particle the parameters $\Lambda_{\underline{\alpha}}$ must satisfy $\Lambda_{+ a} = 0 = \Lambda_{+-} = \Lambda_{+ a b} = \ldots$, that is,  any parameter $\Lambda_{\underline{\alpha}}$ with a lowered $+$ index is zero. The resulting algebra which leaves this choice invariant is [7]
$$
{\cal H}= \{ J_{+i} \ , \ J_{ij} \ , \ S_{+ij} \ , \ S_{i_1i_2i_3} \ , \ S_{+i_1\ldots i_5} \ , \  \ldots  \} \ \ , \ \ i,j,\ldots = 2,\ldots 10.
\eqno(2.2)$$
The generators possessing a lowered $+$ index satisfy the commutation relations
$$
[J_{+i},J_{+j}] = 0 \ , \ [S_{+i_1i_2},S_{+j_1 j_2}] = 0 \ , \ [J_{+i},S_{+j_1 j_2}] = 0 \ , \ \ldots \eqno(2.3)$$
\par
In order to obtain a finite-dimensional unitary representation, these commutation relations imply that all generators with a lowered $+$ index be trivially realized in the representation. The remaining generators obey the $I_c(E_{11})$ commutation relations with indices restricted to $i,j = 2,\ldots,10$. The result is that the subalgebra which preserves the above choice of charges is the Cartan involution invariant subalgebra of $E_9$, denoted ${\cal H} = I_c(E_9)$, given by the generators [7]
$$
J_{ij} \ , \ S_{i_1 i_2 i_3} \ , \ S_{i_1 \ldots i_6} \ , \ S_{i_1 \ldots i_8,j} \ , \ S_{i_1 \ldots i_9,j_1 j_2 j_3} \ , \ \ldots \eqno(2.4)$$
\par
These steps are similar to those of the irreducible massless representations of the Poincare algebra. In this case of finds that the generators $J_{+i}$  
commute and in order  to obtain a finite dimensional irreducible representation one takes a representation in which these act to give zero leaving us to take an irreducible representation of the  algebra  SO(D-2). 
\par
The Dynkin diagram of $E_9$ is given by deleting the first two nodes, which correspond to the $+$ and $-$ directions,  from the Dynkin diagram of $E_{11}$;
$$
\matrix{
& & & & & & & & & & & & & & \bullet  & 11 & & & \cr
& & & & & & & & & & & & & & | & & & & \cr
\oplus& - & \oplus & - & \bullet & - & \bullet & - & \bullet & - & \bullet & - & \bullet & - & \bullet & - & \bullet & - & \bullet \cr
1 & & 2 & & 3 & & 4 & & 5 & & 6 & & 7 & & 8 & & 9 & & 10 \cr
}
$$
Here $\oplus$ indicates that the node has been removed from the Dynkin diagram of $E_{11}$. 
\par
In general the theory in  $D$ dimensions can then be obtained from the $E_{11}$ Dynkin diagram by  deleting node $D$ and analysing the theory with respect to the algebra corresponding to the remaining nodes. Thus to find the theory in eleven dimensions we delete node eleven, that is, the top node and analysis the theory when decomposed into the algebra GL(11), while in $D$ dimensions we decompose into representations of  $A_{D-1}\otimes E_{9-D}$ algebra.  To find the   irreducible representations in $D$ dimensions one carries out the same steps and so for the massless irreducible  representation we end up  deleting nodes one and two, as explained  above, as well as node $D$. For the case of eleven dimensions we should therefore delete nodes one and two to arrive at $E_9$ and  decompose this algebra in terms of representations of  the $A_8$ subgroup of $E_9$ as shown in the Dynkin diagram below
$$
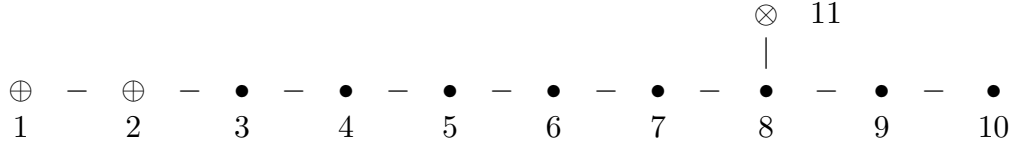
\matrix{
& & & & & & & & & & & & & & \otimes  & 11 & & & \cr
& & & & & & & & & & & & & & | & & & & \cr
\oplus& - & \oplus & - & \bullet & - & \bullet & - & \bullet & - & \bullet & - & \bullet & - & \bullet & - & \bullet & - & \bullet \cr
1 & & 2 & & 3 & & 4 & & 5 & & 6 & & 7 & & 8 & & 9 & & 10 \cr
}
$$
Since  the irreducible representations are representations of $I_c(E_{11})$, rather than $E_{11}$ itself,  we actually decompose the $I_c(E_9)$ subgroup of $E_9$ into representations of $I_c(A_8) = {\rm SO}(9)$. Indeed the  generators of equation (2.4) are listed in this decomposition as from the beginning of our discussion we had in effect deleted node eleven as we are concentrating on the eleven dimensional case. 
\par
The next step is to choose an irreducible representation of $I_c(E_9)$. Such an reducible representation is provided by the Cartan involution odd generators of $E_{9}$ which are given by [7]
$$
T_{ij} = \eta_{ik} K^k{}_k + \eta_{jk} K^k{}_i \ \  , \ \   T_{i_1 i_2 i_3} = R^{j_1 j_2 j_3} \eta_{j_1 i_1} \eta_{j_2 i_2} \eta_{j_3 i_3} + R_{i_1 i_2 i_3} \ , \eqno(2.5) $$
$$
T_{i_1 .. i_6} = R^{j_1 .. j_6} \eta_{j_1 i_1} .. \eta_{j_6 i_6} - R_{i_1 .. i_6} \ , \eqno(2.6) $$
$$
T_{i_1 .. i_8,k} = R^{j_1 .. j_8,l} \eta_{j_1 i_1} .. \eta_{j_8 i_8} \eta_{lk} + R_{i_1 .. i_8,k} \ , \eqno(2.7) $$
$$
T_{i_1 .. i_9,j_1 j_2 j_3} = R^{m_1 .. m_9,l_1 l_2 l_3} \eta_{m_1 i_1} .. \eta_{m_9 i_9} \eta_{l_1 j_1} \eta_{l_2 j_2} \eta_{l_3 j_3} - R_{i_1 .. i_9,j_1 j_2 j_3} \ , \ \ldots  
\eqno(2.8)$$
These generators provide a linear representation of $I_c(E_9)$ because the involution operator $I_c$ is defined to act on the generators $A,B,...$ of a Lie algebra as $I_c(AB) = I_c(A) I_c(B)$ and so $I_c([{\rm even}, {\rm odd}]) = - [{\rm even}, {\rm odd}]$ which guarantees that the commutator will always be a Cartan involution odd generator. The generators of equation (2.8) are the Cartan involution odd generators of  $I_c(E_{11})$ when  restricted to their indices taking only the values $i,j = 2,...,10$. Clearly one could for the massless case take a different irreducible representation but in this paper we will only consider this case. 
\par
As a result we take  our representation to consist of  fields corresponding  to the Cartan involution odd generators 
$$
\{h_{ij}(0) \ , A_{i_1 i_2 i_3}(0) \ , \ A_{i_1 \ldots i_6}(0) \ , \ h_{i_1 \ldots i_8,j}(0) \ , \ \ldots \}, \ \ {\rm with}\ \  \ i,j,... = 2,...,10 \ . 
\eqno(2.9)$$
The value $(0)$ indicates that these fields are before the boost which takes them to be a representation of $I_c(E_{11})\otimes_s l_1$. 
\par
We recognise  $h_{ij}(0)$  and $A_{i_1 i_2 i_3}(0)$ as the degree of freedom of the graviton and three form respectively. There are however, an infinite number of higher level fields.  It was proposed  in [7] that these fields are connected to the gravition and three form by certain duality relations. The purpose of this section  is to investigate these relations in more detail.
\par
The fields of equation (2.9) live in the representation associated to the Cartan involution odd generators and so we consider the quantity
$$
\overline{{\cal V}} = h_{ij} T^{ij} + A_{i_1 i_2 i_3} T^{i_1 i_2 i_3} +  A_{i_1 .. i_6} T^{i_1 .. i_6} + h_{i_1 .. i_8,j} T^{i_1 .. i_8,j} + A_{i_1 .. i_9,j_1j_2 j_3} T^{i_1 .. i_9,j_1j_2j_3} + \ldots 
\eqno(2.10) $$
The transformations of the fields follow from that of the  Cartan involution odd generators under  $I_c(E_9)$. In particular under the  level one  generator $S_{i_1 i_2 i_3}$  of $I_c(E_9)$ we take 
$$
\delta \overline{{\cal V}} = [\Lambda^{i_1 i_2 i_3} S_{i_1 i_2 i_3},\overline{{\cal V}}] + [\Lambda^{i_1 \ldots i_6} S_{i_1 \ldots i_6},\overline{{\cal V}}]
+ [\Lambda^{i_1 \ldots i_6} S_{i_1 \ldots i_6},\overline{{\cal V}}] + [\Lambda^{i_1 \ldots i_8,j} S_{i_1 \ldots i_8,j},\overline{{\cal V}}]
\eqno(2.11)$$
Using the $E_{11}$ algebra to evaluate equation (2.11) we find that
$$
\delta h_{ij} = 18 \, \Lambda_{(i|k_1 k_2|}A_{j)}{}^{k_1 k_2} - 2 \, \eta_{ij} \, \Lambda^{k_1 k_2 k_3} A_{k_1 k_2 k_3} - 5! \, \eta_{ij} \, \Lambda^{k_1 \ldots k_6} A_{k_1 \ldots k_6} - 9 \cdot 5! \, \Lambda^{k_1 \ldots k_5}{}_{(i} A_{j) k_1 \ldots k_5} 
$$
$$
+ 7 \cdot 9 \cdot 10 \cdot 16 \, ( h_{k_1 \ldots k_8,(i|} \Lambda^{k_1 \ldots k_8}{}_{,|j)} + 8 h_{(i| k_1 \ldots k_7,l} \Lambda_{|j)}{}^{k_1 \ldots k_7,l} -  \eta_{ij} \Lambda^{k_1 \ldots k_8,l} h_{k_1 \ldots k_8,l} )   \eqno(2.12) $$
$$
\delta A_{i_1 i_2 i_3}  = - 3 \cdot 2 \, h_{[i_1}{}^k \Lambda_{i_2 i_3]k} + {5! \over 2} \, A_{i_1 i_2 i_3}{}^{k_1 k_2 k_3} \Lambda_{k_1 k_2 k_3} + {5! \over 2} \, \Lambda_{i_1 i_2 i_3 k_1 k_2 k_3} A^{k_1 k_2 k_3} $$
$$
+ {7! \cdot 2 \over 3} \, \Lambda^{k_1 \ldots k_6} (h_{i_1 i_2 i_3 [k_1 \ldots k_5,k_6]} - h_{k_1 \ldots k_6 [i_1 i_2 , i_3]} ) \eqno(2.13) $$
$$
\delta A_{i_1 .. i_6}  =  2 \, \Lambda_{[i_1 i_2 i_3} A_{i_4 i_5 i_6]} + 112 \, h_{i_1 .. i_6 k_1 k_2,k_3} \Lambda^{k_1 k_2 k_3} + 112 \, h_{[i_1 .. i_5 | k_1 k_2 k_3|,i_6]} \Lambda^{k_1 k_2 k_3} $$
$$
+ 12 \, \Lambda_{k[i_1 \ldots i_5} h_{i_6]}{}^{k} \eqno(2.14) $$
$$
\delta h_{i_1 .. i_8,j}  = 2 \, ( \Lambda_{[i_1 i_2 i_3} A_{i_4 .. i_8]j} - \Lambda_{j[i_1 i_2} A_{i_3 .. i_8]} )  - 12 \cdot 8 \cdot 3 \, \Lambda^{k_1 k_2 k_3} ( A_{i_1 .. i_8k_1,k_2k_3 j} + A_{[i_1 .. i_7|jk_1,k_2k_3|i_8]}  ) $$
$$
+ 3 \, \Lambda_{[i_1 \ldots i_6} A_{i_7 i_8] j} - 3 \, \Lambda_{[i_1 \ldots i_6} A_{i_7 i_8 j]}. 
\eqno(2.15)$$
In this section we will only use the $ \Lambda_{i_1 i_2 i_3} $ variations. 
\par
We will now going to postulate duality relations and show they are preserved under the $I_c(E_9)$ symmetry and as a result the number of fields in the representation is radically reduced, indeed, one can take the representation to contain only the graviton and the three form. 
We first propose  a duality relation between the three-form and six-form  which is given by 
$$
E_{i_1 i_2 i_3} = A_{i_1 i_2 i_3} + c \,  \varepsilon_{i_1 i_2 i_3}{}^{j_1 .. j_6} A_{j_1 .. j_6} = 0
 \eqno(2.16)$$
 where  $c$ is a constant which we will fix by requiring that this relation is part of an infinite set of relations that are,  as a collection,  left invariant by 
 $I_c(E_9)$. Varying equation (2.16) under $I_c(E_9)$ we find that 
 $$
\delta E_{i_1 i_2 i_3}  = 2 c \Lambda_{j_1 j_2 j_3}   \varepsilon_{i_1 i_2 i_3}{}^{j_1 .. j_6} (  A_{j_4 j_5 j_6} + {1 \over c} {1 \over 3! 4!} \varepsilon_{j_4 j_5 j_6}{}^{i_1 i_2 i_3 j_1 j_2 j_3} A_{i_1 i_2 i_3 j_1 j_2 j_3} ) + \ldots
\eqno(2.17)$$
where $+\ldots$ denotes the gravity terms. Clearly we will only recover our original relation if 
 $c = \pm {1 \over 12}$ and  we  choose $c = - {1 \over 12}$. For this choice the variation takes the form 
$$
\delta E_{i_1 i_2 i_3}  = {1 \over 6}  \Lambda_{j_1 j_2 j_3}   \varepsilon_{i_1 i_2 i_3}{}^{j_1 .. j_6} E_{j_4 j_5 j_6} - 6 E_{k[i_1} \Lambda^{k}{}_{i_2 i_3]}
\eqno(2.18) $$
where 
$$
E_{i_1 i_2 i_3} = A_{i_1 i_2 i_3} + {1 \over 12} \varepsilon_{i_1 i_2 i_3}{}^{j_1 .. j_6} A_{j_1 .. j_6} = 0 \eqno(2.19) $$
$$
E_{ij} = h_{ij} - {1 \over 4} \varepsilon_i{}^{r_1 .. r_8} h_{r_1 .. r_8,j} = 0.
 \eqno(2.20)$$
\par
Thus we not only recover our original duality relation but find the  new duality relation  of equation (2.20) which relates the graviton and dual graviton. The irreducibility condition $h_{[i_1 .. i_8,j]} = 0$ implies that for this relation to be consistent, the field $h_{ij}$ must be traceless, $h^i{}_i = 0$. With this condition $h_{ij}$  has ${10.9 \over 2}-1= 44$ degrees of freedom as it should. 
\par
We now vary the duality relation of equation (2.20) to find that 
$$
\delta E_{ij} = -  {1 \over 2} {1 \over 3 \cdot 5!} \, \varepsilon_i{}^{r_1 .. r_8}(\Lambda_{r_1 r_2 r_3} \varepsilon_{r_4 .. r_8 j}{}^{k_1 k_2 k_3} E_{k_1 k_2 k_3} - \Lambda_{jr_1 r_2} \varepsilon_{r_3 .. r_8}{}^{k_1 k_2 k_3} E_{k_1 k_2 k_3} )  $$
$$
+ 12 \cdot 2 \cdot 3 \, \Lambda^{k_1 k_2 k_3} \varepsilon_i{}^{r_1 .. r_8}  (E_{r_1 .. r_8k_1,k_2k_3j}  - E_{jr_1 .. r_7k_1,k_2k_3r_8} ) = 0
\eqno(2.21)$$

where $$
E_{i_1 .. i_9,j_1 j_2 j_3} = A_{i_1 .. i_9,j_1 j_2 j_3} + {1 \over 9!} \ \varepsilon_{i_1 .. i_9} A_{j_1 j_2 j_3} = 0.
 \eqno(2.22)$$
Thus the variation contains the  previously-derived duality relation of equation (2.13), and a new duality relation relating the three-form $A_{i_1 i_2 i_3}$ and nine-three form $A_{i_1 \ldots i_9,j_1 j_2 j_3}$.
\par
These duality relations $E_{ij} = 0$, $E_{i_1 i_2 i_3} = 0$, $E_{i_1 \ldots i_9,j_1 j_2 j_3} = 0$ are the first of an infinite tower of duality relations showing that the fields at levels, two, three, four, etc... can be expressed in terms of the fields
$$
h_{ij}, \ \ \ A_{i_1 i_2 i_3}. 
\eqno(2.23)$$
The six form is related to the three form by equation (2.19) and the dual graviton field $h_{i_1\ldots i_9, j}$ is related to the graviton by equation (2.20). All the  higher level fields carry the indices of these fields as well a multiple sets of blocks of nine indices and we can expect that in all the higher level duality relations these indices are carried by $\varepsilon_{i_1 .. i_9} $ 's in a way that is similar to how they appear  in equation (2.22) . These indices correspond to the action of the affine generator of $I_c(E_9)$. As a result it should be possible of show that the complete  infinite set of duality relations are invariant under the $I_c(E_9)$ symmetry. 
\par
Hence we have found that the massless irreducible representation contains the fields of equation (2.23) and as a result it 
contains the  $44 + 84 = 128$ bosonic degrees of freedom of eleven-dimensional supergravity. Rather than write them as in equation (2.23) we can 
write the degrees of freedom in terms of an $E_8$ multiplet, or in our case here,  an $I_C(E_8)=SO(16)$ multiplet, namely 
$$
h_{i'j'} \ , A_{i_1' i_2' i_3'} \ , A_{i_1' \ldots i_6'} \ , \ h_{j'}\equiv {1\over 8!} \epsilon^{i_1' \ldots i'_8} h_{i_1' \ldots i'_8,j'}, \ i',j', \ldots =3,\ldots ,10
 \eqno(2.24)$$
In this paper we will take un-primed indices to range over $i,j,\ldots = 2,\ldots,10$ and primed indices to range over $i',j',\ldots = 3,\ldots,10$. The fields of equation (2.24) have been obtained from those of  equation (2.23) by expressing  $A_{i_1' i_2' 2}$ in terms of $A_{i_1' \ldots i_6'} $ using the duality relation of equation 
(2.16) and also by expressing $h_{2j'} $ in terms of $h_{i_1' \ldots i_8',j'}$ using the duality relation of equation (2.20). The fields of equation (2.24) also give $36+56+28+8=128$ degress of freedom. We note that $h_{i'}{}^{i'}+ h_9{}^9=0$ and as we have not included $h_9{}^9$ we do not take $h_{i'}{}^{i'}$ equal to zero. 
\medskip
{{\bf 3. Reduction of degrees of freedom from the algebra viewpoint}}
\medskip
In the previous section we have seen that the higher level fields can be expressed in terms of the fields $h_{ij}$ and $A_{i_1 i_2 i_3}$ at levels zero  and one  through the existence of duality relations such as those of equations (2.19), (2.20) and (2.22). In this section we will  show that the massless irreducible representation is annihilated by an infinite set  of  generators  and this corresponds to the existence of the infinite set of duality relations.
We begin by considering the  generator of the form
$$
N_{k_1 k_2 k_3} = S_{k_1 k_2 k_3} + c_1 \ \varepsilon_{k_1 k_2 k_3}{}^{r_1 \ldots r_6} S_{r_1 \ldots r_6} 
\eqno(3.1) $$
for a constant $c_1$. Under the action of  this generator the fields transform under ${\Lambda}^{k_1 k_2 k_3} N_{k_1 k_2 k_3}$ as
$$
\delta h_{ij} = 18 {\Lambda}_{(i|k_1 k_2|}A_{j)}{}^{k_1 k_2} - 2 \eta_{ij} {\Lambda}^{k_1 k_2 k_3} A_{k_1 k_2 k_3} $$
$$
- 5! c_1 ( \cdot 3 \cdot 3 \ {\Lambda}^{k_1 k_2 k_3}  \varepsilon_{k_1 k_2 k_3}{}^{r_1 \ldots r_5}{}_{(i} A_{j)}{}_{r_1 .. r_5} + \eta_{ij} {\Lambda}^{k_1 k_2 k_3}  \varepsilon_{k_1 k_2 k_3}{}^{r_1 \ldots r_6} A_{r_1 .. r_6}  ) 
\eqno(3.2) $$
$$
\delta A_{i_1 i_2 i_3}  = - 3 \cdot 2 h_{[i_1}{}^j {\Lambda}_{i_2 i_3]j} + 60 A_{i_1 i_2 i_3}{}^{k_1 k_2 k_3} {\Lambda}_{k_1 k_2 k_3} - {5! \over 2} c_1  \ {\Lambda}^{k_1 k_2 k_3} A^{r_1 r_2 r_3} \varepsilon_{k_1 k_2 k_3 r_1 r_2 r_3}{}^{i_1 i_2 i_3}     $$
$$
- {7! \cdot 2 \over 3} c_1\ {\Lambda}^{k_1 k_2 k_3} ( \varepsilon_{k_1 k_2 k_3}{}_{r_1 \ldots r_6}  A^{r_1 .. r_6}{}_{[i_1 i_2,i_3]}  - \varepsilon_{k_1 k_2 k_3}{}_{r_1 \ldots r_5 j} A_{i_1 i_2 i_3}{}^{r_1 ..r_5,j} )
 \eqno(3.3)$$
$$
\delta A^{i_1 \ldots i_6} = 2 {\Lambda}^{[i_1 i_2 i_3} A^{i_4 i_5 i_6]} + 112 h^{i_1 .. i_6 k_1 k_2,k_3} {\Lambda}_{k_1 k_2 k_3} + 112 h^{[i_1 .. i_5 | k_1 k_2 k_3|,i_6]} \tilde{\Lambda}_{k_1 k_2 k_3} $$
$$
+ 12 \ c_1 \ {\Lambda}^{k_1 k_2 k_3} \varepsilon_{k_1 k_2 k_3 j}{}^{[i_1 \ldots i_5} h^{i_6]j}  $$
$$
- 3 \cdot 7! c_1 \, {\Lambda}^{k_1 k_2 k_3}  \varepsilon_{k_1 k_2 k_3}{}_{r_1 \ldots r_6} A^{[i_1 .. i_4 | [r_1 .. r_5 ,r_6] | i_5 i_6]}     $$
$$
- 15 \cdot 7! c_1 \, {\Lambda}^{k_1 k_2 k_3} \varepsilon_{k_1 k_2 k_3}{}_{r_1 \ldots r_4 r_5 r_6}  A^{[r_1 .. r_4 | [i_1 .. i_5, i_6] | r_5 r_6]}  $$
$$
- 4 \cdot 7! c_1 {\Lambda}^{k_1 k_2 k_3} \varepsilon_{k_1 k_2 k_3}{}_{r_1 \ldots r_6} A^{r_1 .. r_6 [i_1 i_2 i_3,i_4 i_5 i_6]} \eqno(3.4) $$
$$
\delta h_{i_1 \ldots i_8,j} = 2 ( {\Lambda}_{[i_1 i_2 i_3} A_{i_4 .. i_8]j} - \tilde{\Lambda}_{j[i_1 i_2} A_{i_3 .. i_8]} ) - 12 \cdot 8 \cdot 3 {\Lambda}^{k_1 k_2 k_3} ( A_{i_1 .. i_8k_1,k_2k_3 j} + A_{[i_1 .. i_7|jk_1,k_2k_3|i_8]} ) $$
$$
+ 3  c_1 \ {\Lambda}^{k_1 k_2 k_3} \varepsilon_{k_1 k_2 k_3}{}^{[i_1 \ldots i_6} A^{i_7 i_8] j} - 3  c_1 \ {\Lambda}^{k_1 k_2 k_3} \varepsilon_{k_1 k_2 k_3}{}^{[i_1 \ldots i_6} A^{i_7 i_8 j]} 
\eqno(3.5) $$
On setting $c_1 = - {1 \over 3 \cdot 5!}$ in the generator of equation (3.1) becomes 
$$
N_{k_1 k_2 k_3} = S_{k_1 k_2 k_3} - {1 \over 3 \cdot 5!} \ \varepsilon_{k_1 k_2 k_3}{}^{r_1 \ldots r_6} S_{r_1 \ldots r_6} \ \ , 
\eqno(3.6) $$
With this choice the variations of equations  (3.2) to (3.5) can be expressed in terms of the duality relations defined in equations (2.19) (2.20) and (2.22) and  become equal to zero
$$
\delta h_{ij} =  18 {\Lambda}_{(i}{}^{ k_1 k_2} E_{j) k_1 k_2}  - 2 \eta_{ij} {\Lambda}^{k_1 k_2 k_3} E_{k_1 k_2 k_3}  = 0 \eqno(3.7)$$
$$
\delta A_{i_1 i_2 i_3} = {1 \over 3!} \varepsilon_{i_1 i_2 i_3 k_1 k_2 k_3}{}_{r_1 r_2 r_3} {\Lambda}^{k_1 k_2 k_3} E^{r_1 r_2 r_3} - 6  {\Lambda}_{[i_1 i_2}{}^{k}  E_{i_3]k} = 0 \eqno(3.8) $$
$$
\delta A_{i_1 \ldots i_6}  =  {1 \over 30}  {\Lambda}^{k_1 k_2 k_3} \varepsilon_{k_1 k_2 k_3[i_1 \ldots i_5|}{}^j E_{|i_6]}{}_{j} + 28 {\Lambda}_{k_1 k_2 k_3} \varepsilon^{k_1 k_2 k_3}{}^{r_1 \ldots r_6} E_{r_1 .. r_6 [i_1 i_2 i_3,i_4 i_5 i_6]} \eqno(3.9)  $$
$$
-  140 {\Lambda}_{k_1 k_2 k_3} \varepsilon^{k_1 k_2 k_3 r_1 \ldots r_4 r_5 r_6}  E_{i_1 .. i_6 [r_1 r_2 r_3 , r_4 r_5 r_6]} = 0 
\eqno(3.10)$$
\par
In a similar fashion one can identify other generators that annihilate the massless irreducible representation. Indeed one finds that the  generator 
$$
N_{ij} = J_{ij} + {2 \over 8!} \varepsilon_{[i}{}^{r_1 \ldots r_8} S_{|r_1 \ldots r_8|,j]} 
\eqno(3.11) $$
The variation of the graviton under a transformation of this generator with parameter $ {\Lambda}^{ij}$ gives 
$$
\delta h_{ij} = 2 {\Lambda}_{i}{}^{k} E_{k j} + 2 {\Lambda}_j{}^{k} E_{ki} =0 , 
\eqno(3.12)$$
It would seem inevitable that one has an infinite number of generators that annihilate the entire representation and we will take this to be the case. At the next level we would expect to find the generator 
$$
\tilde N_{i_1i_2i_3} \equiv  S_{i_1i_2i_3} +{1\over 9!} \epsilon^{j_1\ldots j_9} S_{j_1\ldots j_9 ,} {}^{i_1i_2i_3}
\eqno(3.13)$$
where we have used the value of the coefficient that we will find below. The pattern of the higher level generators that annihilate the representation is apparent. To a given generator one adds with a suitable coefficient another generator which possess an additional block  of nine anti-symmetric indices  which are saturate  with the $\epsilon$ symbol. 
\par
The set of all generators that annihilate the representation must form a subalgebra, denoted $I$, of $I_c(E_9)$ and we will now show that this is true for such  lowest level  generators. Indeed we find that 
$$
[N^{i_1 i_2 i_3},N_{j_1 j_2 j_3}] = 2 N^{i_1 i_2 i_3}{}_{j_1 j_2 j_3} - 9 \cdot 4 \delta^{[i_1 i_2}_{[j_1 j_2} N^{i_3]}{}_{j_3]} 
 - {1 \over 2} \varepsilon^{i_1 i_2 i_3}{}_{j_1 j_2 j_3 k_1 k_2 k_3}  \tilde N^{k_1 k_2 k_3} 
\eqno(3.14) $$
where 
$$N_{i_1 ... i_6} = S_{i_1 .. i_6} + {1 \over 12} \varepsilon_{i_1 .. i_6}{}^{k_1 k_2 k_3} S_{k_1 k_2 k_3}= {1\over 12} \varepsilon_{i_1 .. i_6}{}^{k_1 k_2 k_3} N_{k_1k_2k_3}
\eqno(3.15)$$ 
\par
The generators that annihilate the irreducible representation also form an ideal.  We recall that an ideal $I$ of a Lie algebra $G$ is a subalgebra consisting  of elements $X \in I$ such that such that $[X,Y] \in I$ for all $Y \in G$. One finds that  
$$
[S^{i_1 i_2 i_3},N_{j_1 j_2 j_3}] = - {1 \over 6} \varepsilon^{i_1 i_2 i_3}{}_{j_1 j_2 j_3 k_1 k_2 k_3} N^{k_1 k_2 k_3} - 18 \delta^{[i_1 i_2}_{[j_1 j_2} N^{i_3]}{}_{j_3]} 
\eqno(3.16) $$
and that 
$$
[S_{i_1 i_2 i_3},N_{jk}] = 3 \eta_{[i_1 |j} \tilde N_{k| i_2 i_3]} + 3 \eta_{[i_1 |j} N_{k| i_2 i_3]} 
\eqno(3.17)$$
Since the commutators of the generator $S^{i_1 i_2 i_3}$ lead to  the whole of $I_c(E_9)$ it follows that the equations (3.16) and (3.17) together with their higher level analogues, which we have not shown,  imply that  the generators in the subalgebra $I$ are  an ideal of $I_c(E_9)$. 
 Starting from the level zero field $h_{ij}(0)$ of the massless irreducible representation we can find all fields  in the representation  by the action of $S^{i_1 i_2 i_3}$. Hence if the generators  which annihilate the massless irreducible representation  form an ideal in $I_c(E_9)$ it follows  that  if all the elements of $I$ annihilate the lowest field, the graviton, then they will annihilate all fields in the representation.  
\par
Given a Lie algebra $G$ (group) which contains an ideal subalgebra (subgroup) $I$ then the coset $G\over I$ is also a Lie algebra (group). If $A_1,  A_2$ belong to $G$ the corresponding  equivalence relation is  $A_1\sim A_2$ means $A_1=  A_2+i$ for some $i\in I$. Indeed $I_c(E_9)\over I$ is a Lie algebra. It is clear from the form of the generators in the ideal that all the generators in $I_c(E_9)$ are related to the generators 
$J_{i_1i_2} $ and $S_{i_1i_2i_3}$ by the equivalence relations. Hence $I_c(E_9)\over I$ contains just these two generators. These generators  contain $36+84=120$ generators and, as we will show in the next next section,  they generate S0(16). As a result 
$$
{I_c(E_9)\over I}= SO(16)
\eqno(3.18)$$
 \par
 In this section we have seen that there is a  radical reduction in the number of states of the massless irreducible representation can be traced to the existence of a ideal in $I_c(E_9)$. It contains only the graviton and three form.   Indeed due to equation (3.18) these states  belong to an irreducible representation of the much smaller algebra SO(16). The situation is a bit similar to the existence of highest weight states in representations of the Virasoro algebra and the corresponding reduction in the number of  states in the representation. 
 \medskip
{{\bf 4. The bosonic states viewed as a spinor}}
\medskip
In the section two we found that although the massless irreducible representation of the little group $I_c(E_9)$ contained,  at first sight, an  infinite number of states  it was subject to an infinite set of duality relations that reduced the number of degrees of freedom it contained to 128. These can be listed as in equation (2.24),  or as an $I_c(E_8)={\rm SO}(16)$ multiplet,  in equation (2.24). The SO(16) algebra has two 128 dimensional irreducible representations both of which are spinors. The number of components  of a spinor in sixteen  dimensions is $2^{{16\over 2}}= 256$. However, in sixteen, effectively Euclidean, dimensions we can have Majorana Weyl spinors which have just  128 component. Hence it must be that our 128 bosonic states belong to a spinor representation of SO(16). As $E_8$ contains the gravity line consisting of the nodes three to ten it contains the SL(9)  algebra and as a result  $I_c(E_8)={\rm SO}(16)$ contains the SO(9) which acts on the spacetime time coordinates. The purpose of this section is to understand how the bosonic states can be assembled into this  spinor representation. 
\par
We found that for the massless representation the little algebra  is $I_c(E_9)$ which arises when we delete  nodes one and two  in the $E_{11}$ Dynkin diagram.   Further decomposing  the  $I_c(E_9)$  representations  into  those of $I_c(E_8)$ corresponds  to also deleting  node three in addition to  nodes one and two in   the  $E_{11}$ Dynkin diagram  as shown below 
$$
\matrix{
& & & & & & & & & & & & & & \bullet  & 11 & & & \cr
& & & & & & & & & & & & & & | & & & & \cr
\oplus& - & \oplus & - & \otimes & - & \bullet & - & \bullet & - & \bullet & - & \bullet & - & \bullet & - & \bullet & - & \bullet \cr
1 & & 2 & & 3 & & 4 & & 5 & & 6 & & 7 & & 8 & & 9 & & 10 \cr
}
$$
\par 
The eleven dimensional theory appears when  we deleted node eleven and  decompose with respect to the remaining $A_{10}$ algebra. In the context of  the massless irreducible  representation, which has little group $I_c(E_9)$,  this implies that we should decompose with respect to  $I_c(A_8)={\rm  SO}(9)$. Indeed  the fields of equation  (2.9), and so equation (2.23),  appear as representations of  $I_c(A_8)={\rm  SO}(9)$.
However, we wish to study the  massless irreducible representation from the viewpoint of  $I_c(E_8)={\rm SO}(16)$  which is the algebra that appears in the above $E_{11}$ Dynkin diagram. As such we have in effect to undo the deletion of node eleven. 
 To better understand the manner in which bosonic fields belong to the 128 dimensional spinor representation   we consider the  fields from the viewpoint of the  ${\rm SO}(8) \times {\rm SO}(8)$ subalgebra of 
${\rm  SO}(16)$. 
\par 
The first step is to identify the ${\rm SO}(8) \times {\rm SO}(8)$ generators amongst those of  $I_c(E_8)$. The latter generators   arise in $E_{11}$  as 
$$
\{ J_{i'j'} \ , \ S_{i_1' i_2' i_3'} \ , \ S_{i_1' \ldots i_6'} \ , \ S_{i_1' \ldots i_8',j'} \}, \ \  {\rm with}\ \  i'. j',\ldots 3,\ldots , 10
\eqno(4.1) $$
These Cartan involution invariant generators appear  in equations (2.5-2.8) if  we change the signs between the two generators from plus to minus and visa-versa. 
In equation (4.1)  one finds $28 + 56 + 28 + 8 = 120$ generators. The $J_{i'j'}$ generate an ${\rm SO}(8)$ This subalgebra  is the $I_c(A_7)$ that appears as   the gravity line consisting of nodes four to ten of the $E_{11}$ Dynkin diagram.  As such this SO(8) is just part of the familiar gravity line symmetries that are foremost in E theory discussions and act on the spacetime coordinates. The ${\rm SO}(8) \times {\rm SO}(8)$ subalgebra consists of  the generators 
$$
J^{\pm}_{i'j'} = J_{i'j'} \pm 2 \hat{J}_{i'j'} , \quad \quad {\rm where}\quad \quad  
\hat{J}_{i'j'} = {1 \over 6!} \varepsilon_{i'j'}{}^{k_1' \ldots k_6'} S_{k_1' \ldots k_6'}. 
\eqno(4.2)$$
Using the well known commutators of the $E_{11}$ algebra one finds that they obey the commutation relations 
$$
[J^+{}^{i'j'},J^-_{k'l'}] = 0 \ , 
\eqno(4.3) $$
$$
[J^+{}^{i'j'},J^+_{k'l'}] = - 8 \delta^{[i'}{}_{[k'} J^+{}^{j']}{}_{l']}  \ , \ [J^-{}^{i'j'},J^-_{k'l'}] = - 8 \delta^{[i'}{}_{[k'} J^-{}^{j']}{}_{l']}. 
\eqno(4.4)$$
which are indeed those of ${\rm SO}(8) \times {\rm SO}(8)$. We note that the gravity line SO(8) discussed above arises as  the algebra which is the diagonal subalgebra of ${\rm SO}(8) \times {\rm SO}(8)$. We note that we have three different SO(8) algebras. 
\par
We will now decompose  the  spinor representation of SO(16), which contains the ${\rm 128}$ degrees of freedom, into  representations of  ${\rm SO}(8) \times {\rm SO}(8)$. To begin with rather than consider the fields we will consider the corresponding Cartan involution odd generators of equation (2.5-2.8). 
While these transform into each other in the usual way under the SO(8) rotations generated by $J_{i'j'} $,  under the $S_{i_1' \ldots i_6'}$ generator we find that 
$$
[S_{i_1' \ldots i_6'}, T^{j'k'}] = - 6 T_{[i_1' \ldots i_5'}{}^{j'} \delta^{k'}{}_{i_6']} - 6 T_{[i_1' \ldots i_5'}{}^{k'} \delta^{j'}{}_{i_6']}, 
\eqno(4.5) $$
$$
[S_{i_1' \ldots i_6'}, T^{j_1' j_2' j_3'}] = 60 T_{[i_1' i_2' i_3'} \delta^{j_1' j_2' j_3'}_{i_4' i_5' i_6']} + 3 T_{i_1' \ldots i_6'}{}^{[j_1' j_2' , j_3']} , 
\eqno(4.6) $$
$$
[S_{i_1' \ldots i_6'}, T^{j_1' \ldots j_6'}] = - 120 T^{k'}{}_{k'} \delta^{j_1' \ldots j_6'}_{i_1' \ldots i_6'} + 1080 T_{[i_1'}{}^{[j_1'} \delta^{j_2' \ldots j_6']}_{i_2' \ldots i_6']}, 
\eqno(4.7) $$
$$
[S_{i_1' \ldots i_6'}, T^{j_1' \ldots j_8',k'}] = 3360 (T^{[j_1' j_2' j_3'} \delta^{j_4' \ldots j_8'] k'}_{i_1' \ldots i_5' i_6'} - T^{[j_1' j_2' |k'|} \delta^{j_3' \ldots j_8']}_{i_1' \ldots i_6'}), 
\eqno(4.8)$$
Examining these equations we notice that $ T^{j'k'}$ and $\tilde {T}_{i'j'} \equiv  {1 \over 6!} \varepsilon_{i'j'}{}^{k_1' \ldots k_6'} T_{k_1' \ldots k_6'} $ transform into each other as do $T^{j_1' j_2' j_3'}$ and $T^{i'} \equiv {1 \over 8!} \varepsilon_{k_1' \ldots k_8'} T^{k_1' \ldots k_8',i'}$. 
\par
Motivated by the result just above we  define the combination 
$$
\hat T{}_{i'}{}^{j'} = T_{i'}{}^{j'} + 2 \tilde {T}_{i'}{}^{j'}-{1\over 6}\delta  _{i'}^{j'} T_{k'}{}^{k'}
\eqno(4.9)$$
which contains $8.8=64$ degrees of freedom as $T_{i'j'}$ and $ \tilde{T}_{i'j'}$ are symmetric and antisymmetric respectively.
Under ${\rm SO}(8) \times {\rm SO}(8)$ the objects $\hat T_{i'j'} $  transforms as  
$$
[J^{+i' j'},\hat T{}_{k'}{}_{l'}] = - 4 \delta^{[i'}{}_{k'} \hat T{}^{j']}{}_{l'}, \ \ 
[J^{-i' j'},\hat T{}_{k'}{}_{l'}] = - 4 \delta^{[i'}{}_{l'} \hat T{}_{k'}{}^{j']} 
\eqno(4.10)$$
We recognise that the first ${\rm SO}(8)$ transforms the first index on $\hat T{}_{k'}{}_{l'}$ as a vector and the second ${\rm SO}(8)$
transforms the second  index on $,\hat T{}_{k'}{}_{l'}$ as a vector.  Thus we recognise that $\hat T{}_{k'}{}_{l'}$ transforms as the $(8_v,8_v)$  representation of ${\rm SO}(8) \times {\rm SO}(8)$ where $8_v$ is the vector representation of SO(8). 
\par
We now turn our attention to the objects $T^{j_1' j_2' j_3'}$ and $T^{i'} $ which have $56$ and $8$ degrees of freedom respectively. Under ${\rm SO}(8) \times {\rm SO}(8)$ they transform as 
$$
[J_{i'j'}^{\pm},T_{k'}] = 2 T_{[i'} \delta_{j'] k'} \mp{1 \over 2} T_{i'j'k'} 
\eqno(4.11) $$
and 
$$
[J^{\pm}{}_{i'j'},T_{k'_1 k'_2 k'_3}] = 6 T_{[k'_1 k'_2 | [i'} \delta_{j'] | k'_3]} \mp {1 \over 6} \varepsilon_{i'j' k'_1 k'_2 k'_3 l'_1 l'_2 l'_3} T^{l'_1 l'_2 l'_3} 
\pm 12\delta _{i'j'}^{[k_1' k_2'} T^{k_3']}
\eqno(4.12) $$
Hence  $T_{k'}$ and $T_{i_1' i_2' i_3'}$ and  form the $56 + 8 = 64$ dimensional  representation of  ${\rm SO}(8) \times {\rm SO}(8)$. 
\par
We will now identify what representation  this is. We observe that the eight dimensional gamma matrices in Euclidean space obey the equations 
$$
\gamma_{i'j'} \gamma_{k'} = \gamma_{i'j'k'}+ 2\delta_{k' [j' } \gamma _{i']}
\eqno(4.13) $$ 
and 
$$
\ \ \   \gamma_{i'j'}\gamma^{k'_1 k'_2 k'_3}=6 \delta_{[j'}{}^{[k'_1} \gamma_{i']}{}^{k'_2 k'_3]} + \gamma_{i'j'}{}^{k'_1 k'_2 k'_3} -6\delta _{i'j'}^{[k_1' k_2'} \gamma^{k_3']}
$$
$$
= 
6 \delta_{[j'}{}^{[k'_1} \gamma_{i']}{}^{k'_2 k'_3]} +{1 \over 3!} \varepsilon_{i'j'}{}^{k'_1 k'_2 k'_3 l'_1 l'_2 l'_3} \gamma_{l'_1 l'_2 l'_3}\gamma_9 -6\delta _{i'j'}^{[k_1' k_2'} \gamma^{k_3']}
\eqno(4.14) $$ 
where 
$$
\gamma_9\equiv \gamma^{1\ldots 8}
\eqno(4.15)$$
 Hence if we were to make the identifications  
$$
T_{k'} = \gamma_{k'} \ ,  \  T_{i'j'k'} = - { 2} \gamma_{i'j'k'} 
\eqno(4.16) $$ 
 then multiplication on the left by $  \gamma^{i'j'} $ has the same result as $J_{i'j'}^{+}$ in  the commutators of equations (4.11) and (4.12) provided $\gamma_9$ takes the value $-1$. Similarly one can verify that right multiplication by $ - \gamma^{i'j'} $ has the same result as $J_{i'j'}^{-}$ in  the commutators of equations (4.11) and (4.12). 
 \par
 As a result we can think of $T^{j_1' j_2' j_3'}$ and $T^{i'} $ as belonging to a bi-spinor which is Weyl projected. A bi-spinor takes the form 
 $$
 \Gamma = c_0 I+ c_{i'} \gamma^{i'}+ \ldots +c_{i'_1\ldots i'_8} \gamma^{i'_1\ldots i'_8} 
 \eqno(4.17)$$
 However, if we require $\gamma_9 \Gamma=\Gamma$ then we find that $c_0={1\over 8!} \epsilon^{i'_1\ldots i'_8}c_{i'_1\ldots i'_8}$ etc and we can take the bi-spinor to be of the form 
 $$
 \Gamma = (1+\gamma_9)(c_0 I+ c_{i'} \gamma^{i'}+ \ldots +c_{i'_1\ldots i'_4} \gamma^{i'_1\ldots i'_4} )
 \eqno(4.18)$$
 However, we can also demand that $\Gamma\gamma_9= -\Gamma$ then find that $\Gamma $ takes the form 
 $$
 \Gamma =  (1+\gamma_9)(c_{i'} \gamma^{i'}+c_{i'_1\ldots i'_3} \gamma^{i'_1\ldots i'_3} )
 \eqno(4.19)$$
Thus we find the representation carried by $T^{j'_1 j'_2 j'_3}$ and $T^{i'} $ is a bispinor with the above values of Weyl projection. The coefficients $c_{i'}$ and $c_{i'_1 i'_2 i'_3}$ correspond to the fields 
$h_{i'}\equiv {1\over 8!}\epsilon^{j_1\ldots j_8}h_{j_1\ldots j_8,i}$ and $A_{i'_1\ldots i'_3}$ of equation (2.24). It we denote the eight dimensional Majorana Weyl spinor representation of SO(8) with Weyl projection $+1$, that is $\gamma_9\epsilon = \epsilon$,   by $8_c$ and the one with Weyl projection $-1$  by $8_s$ then the generators $T^{j_1' j_2' j_3'}$ and $T^{i'} $ belong to the $(8_c ,  8_s)$ representation of ${\rm SO}(8) \times {\rm SO}(8)$.
\par
 Hence we find that the  the spinor representation of $SO(16)$  which contains the 128 bosonic degrees of freedom of the massless irreducible  representation decomposes into the 
$$
128= (8_v, 8_v) \oplus (8_c ,  8_s) 
\eqno(4.20)$$
representations of ${\rm SO}(8) \times {\rm SO}(8)$. The $(8_v, 8_v)$ contains the fields $h_{i'j'}$ and $A_{i'_1\ldots i'_6} $ of equation (2.24) while the  $(8_c ,  8_s)$ contains the $A_{i'_1i'_2,i'_3}$ and $h_{i'_1 \ldots i'_8, k'}$ fields. The duality equations relate these two representations. The higher level  fields are related by duality equations  to the fields of these two representations and so these duality relations  can be thought of arising from  the action of the affine operator that takes $I_c(E_8)$ to $I_c(E_9)$. 
\par
Rather than discuss the transformations of the Cartan involution odd generators we can consider the transformations of the corresponding fields 
which we will now derive. The Cartan form containing these fields can be written as 
$$
\tilde{{\cal V}} = \hat A{}_{i_1' i_2'} \hat T{}^{i_1' i_2'} + A_{i'} T^{i'} + A_{i_1' i_2' i_3'} T^{i_1' i_2' i_3'} 
 \eqno(4.21)$$
In terms of our original generators this takes the form  
$$
\tilde{{\cal V}} = [(\hat A{}_{i_1' i_2'} - {1 \over 6} \hat A{}^{k'}{}_{k'} \delta_{i_1' i_2'})]T^{i_1' i_2'} + A_{i_1' i_2' i_3'} T^{i_1' i_2' i_3'} + {2 \over 6!}  \varepsilon_{i_1' \ldots i_6'}{}^{k_1' k_2'} \hat A{}_{k_1' k_2'}  T^{i_1' \ldots i_6'}
$$
$$
+ {1 \over 8!} \varepsilon_{i_1' \ldots i_8'} A_{k'} T^{i_1' \ldots i_8',k'}
 \eqno(4.22)$$
Using the commutation relations (4.10), (4.11), (4.12) that

$$
\delta \tilde{{\cal V}} = [\Lambda^{\pm}{}_{i_1' i_2'} J^{\pm}{}^{i_1' i_2'},\tilde{{\cal V}}]
\eqno(4.23)$$

$$
\delta_{\Lambda^+} \hat A{}_{i_1' i_2'} = 4 \Lambda^{+}{}_{i_1' k'} \hat A{}^{k'}{}_{i_2'} \ \ , \ \ \delta_{\Lambda^-} \hat A{}_{i_1' i_2'} = - 4 \hat A{}_{i_1' k'} \Lambda^-{}^{k'}{}_{i_2'} \ \ , $$
$$
\delta_{\Lambda^{\pm}} A_{i'} = 2 \Lambda^{\pm}{}_{i' k'} \hat A^{k'} \pm 12 \Lambda^{+}{}^{k_1' k_2'} A_{k_1' k_2' i'}
\eqno(4.24)$$

$$
\delta_{\Lambda^\pm} A_{i_1' i_2' i_3'} = \mp {1 \over 2} \Lambda^{\pm}{}_{[i_1' i_2'} A_{i_3']} + 6 \Lambda^{\pm}{}_{[i_1' |k'} A^{k'}{}_{|i_2' i_3']} \pm {1 \over 6} \varepsilon_{i_1' i_2' i_3' j_1' j_2' k_1' k_2' k_3'} \Lambda^{\pm}{}^{j_1' j_2'} A^{k_1' k_2' k_3'}
\eqno(4.25)$$
\par
It will also be instructive to decompose the 120 dimensional adjoint representation of $I_c(E_8)={\rm SO}(16)$, whose generators are given in equation equation (4.1),  into representations of terms of  ${\rm SO}(8) \times {\rm SO}(8)$. The generators not included in  ${\rm SO}(8) \times {\rm SO}(8)$ are  $S_{k'}\equiv {1\over 8!} \epsilon^{j'_1\ldots j'_8} S_{j'_1\ldots j'_8, k'}$ and $S_{k'_1 k'_2 k'_3}$            and their commutation relations with the ${\rm SO}(8) \times {\rm SO}(8)$ generators are given by 
$$
[J_{i'j'}^{\pm},S_{k'}] = 2S _{[i'} \delta_{j'] k'} \mp{1 \over 2} S_{i'j'k'} 
\eqno(4.26) $$
and 
$$
[J^{\pm}{}^{i'j'},S_{k'_1 k'_2 k'_3}] = 6  S_{[k'_1 k'_2}{}^{[i'} \delta_{k'_3]}{}^{j']} \pm {1 \over 6}  \varepsilon^{i'j'}{}_{k'_1 k'_2 k'_3 l'_1 l'_2 l'_3} S^{l'_1 l'_2 l'_3}
\pm 12  \delta^{i'j'}_{[k_1' k_2'} S_{k_3']}
\eqno(4.27) $$
Comparing with the commutators of equations (4.11) and (4.12) we see that $S_{k'}$ and $S_{k'_1 k'_2 k'_3}$ have the same commutators with 
${\rm SO}(8) \times {\rm SO}(8)$ as $T_{k'}$ and $T_{k'_1 k'_2 k'_3}$ except for an opposite  sign in the second term on the right-hand side of equation (4.27). Examining the gamma matrix algebra of equations (4.13) and (4.14) we see that we can identify  $S_{k'}$ and $S_{k'_1 k'_2 k'_3}$ with 
 $\gamma_{k'}$ and $-2\gamma_{k'_1 k'_2 k'_3}$ provided we take $\gamma_9$ in equation (4.14)  to take the value $+1$. Hence these generators belong to the 
 $(8_s,8_c)$  representation of ${\rm SO}(8) \times {\rm SO}(8)$. Hence the 120 dimensional adjoint representation of SO(16) consists of the 
$$
120= (28,1)\oplus (1,28)\oplus  (8_s ,  8_c) 
\eqno(4.28)$$
representation of ${\rm SO}(8) \times {\rm SO}(8)$. The $(28,1)\oplus (1,28)$ contain the adjoint representation of ${\rm SO}(8) \times {\rm SO}(8)$. 
We note that the decompositions we have found are not quite the same as those one finds in certain books. 


\medskip
{{\bf 5. Spinors of $I_c(E_9)$ decomposed into representations of  ${\rm SO}(8) \times {\rm SO}(8)$}}
\medskip
Having decomposed the 128 dimensional SO(16) spinor representation that contains the bosonic degrees of freedom  in terms of representation of 
${\rm SO}(8) \times {\rm SO}(8)$ it will be educational to also analyse the 128 dimensional spinor representation to which the fermionic degrees of freedom  belong. Long ago it was shown that the fermionic degrees of freedom appear in maximal supergravity in $D$ dimensions as a linear representation of the Cartan involution invariant subgroup of the duality group $E_{11-D}$, for example in four dimensions this group is $I_c(E_7)=SU(8)$ [16]. It was therefore natural to take the fermions in $E$ theory to be a linear representation of the Cartan involution invariant subalgebra of $I_c(E_{11})$. In fact fermions were first introduced [17-20]  in this way in the context of the $E_{10}$ theory and subsequently  [21] in the $E_{11}$ theory. The key to these constructions was the realisation that $I_c(E_{10})$ and  $I_c(E_{11})$ admit highly unfaithful representations. In particular it was shown that  $I_c(E_{11})$ has a representation based on a spinor of SO(1,10), $\epsilon_\alpha$ in which the low lowest level generators take the form [2]
$$
J_{a_1a_2}\epsilon_\alpha=-{1\over 2} ( \gamma_{a_1a_2})_\alpha {}^\beta\epsilon_\beta, \ \ 
S_{a_1a_2a_3}\epsilon_\alpha={1\over 2} ( \gamma_{a_1a_2a_3})_\alpha {}^\beta\epsilon_\beta, \ \ 
$$
$$
S_{a_1\ldots a_6}\epsilon_\alpha=-{1\over 4} ( \gamma_{a_1\ldots a_6})_\alpha {}^\beta\epsilon_\beta , \ \ 
S_{a_1\ldots a_8,}{}^{ b}= \delta^b_{[a_1}\epsilon_{a_2\ldots a_8] c_1c_2}\gamma^{c_1c_2} , \ldots 
\eqno(5.1)$$
The generators at higher levels can be found by substituting the above actions into the $I_c(E_{11})$ commutation relations. Clearly, some parts of the higher level generators are trivially realised and so the representation is unfaithful. 
\par
We will be interested in the fermionic degrees of freedom from the viewpoint of the irreducible representations of $I_c(E_{11})\otimes _s l_1$ and in particular as a representation of  $I_c(E_{9})$. The  $I_c(E_{9})$ transformations of the spinor $\epsilon_\alpha$ are given by equation (5.1) with the indices restricted to take the values $2,\ldots , 10$. It is straight forward to  show  that the representation,   $\epsilon$ is annihilated by precisely the same generators of equation (3.6) and (3.11), that is,  
$$
N_{i_1i_2i_3} \epsilon=0= N_{i_1i_2} \epsilon
\eqno(5.2)$$
It must be true that there are an infinite number of such generators that also annihilate $\epsilon$. Thus the spinor $\epsilon$ is annihilated by the same generators as the representation that contains the bosonic degrees of freedom  and so they form the  same ideal $I$. As such the spinor is really a representation of ${K(E_9)\over I}=SO(16)$. 
\par
The generalisation of the unfaithful representation of equation (5.1)  to the gravitino which was first done in the context of $E_{10}$ [17-20] and the result for $E_{11}$ [21] is 
$$
J_{a_1 a_2} \psi_{b} = - {1 \over 2} \gamma_{a_1 a_2} \psi_{b} - 2 \eta_{b [a_1} \psi_{a_2]}$$
$$
S_{a_1 a_2 a_3} \psi_{b} = \lambda \{ {1 \over 2} \gamma_{a_1 a_2 a_3} \psi_{b} - \gamma_{b [a_1 a_2} \psi_{a_3]} + 4 \eta_{b [a_1} \gamma_{a_2} \psi_{a_3]} \} $$
$$
S_{a_1 \ldots a_6} \psi_{b} = - { 1 \over 4} \gamma_{a_1 \ldots a_6} \psi_{b} - 2 \gamma_{b [a_1 \ldots a_5} \psi_{a_6]} + 5 \eta_{b [ a_1} \gamma_{a_2 \ldots a_5} \psi_{a_6]} $$
$$
S_{a_1 \ldots a_8 , c} \psi_b = \lambda \{ \gamma_{b [a_1 \ldots a_8} \psi_{c]} - \gamma_{b a_1 \ldots a_8} \psi_{c}
+ 8 ( \eta_{b [c} \gamma_{a_1 \ldots a_7} \psi_{a_8]} - \eta_{b c} \gamma_{[a_1 \ldots a_7} \psi_{a_8]}) $$
$$ 
- 2 \eta_{c [a_1} \gamma_{a_2 \ldots a_8]} \psi_{b} - 28 \gamma_{b [a_1 \ldots a_6} \psi_{a_7} \eta_{a_8] c} \} \eqno(5.3) $$
We can take the constant  $\lambda$ to take the values $\lambda  = - 1$ or $\lambda  = 1$ and still have a representation of $I_c(E_{11})$. This reflects the way the generator  $S_{a_1 a_2 a_3}$ appears in the algebra resulting from its level one character. In reference [21] we took 
 $\lambda  = 1$ but in this paper we will find that the other sign is better. 
\par
When  the indices in equation (5.2) take the range $3$ to $10$ we can interpret this equation as containing the representation that contains the fermionic degrees of freedom as they occur in the little algebra $I_c(E_{9})$. 
\par
In this section we will decompose the spinor representation of equation (5.2) into those of ${\rm SO}(8) \times {\rm SO}(8)$  which is generated by 
$J^\pm_{i'j'} $ defined in equation (4.2), but we will begin with the representation of equation (5.1). We find that 
$$
J^\pm_{i'j'} \epsilon=  - {1 \over 2} \gamma_{i'j'} (1\mp \gamma_9) \epsilon
\eqno(5.4)$$
where $\gamma_9=\gamma_1\ldots \gamma_8$. 
Defining $\epsilon_\pm= {1\over 2} (I\pm\gamma_9)\epsilon$ we can rewrite this equation as 
$$
J^+_{i'j'} \epsilon_+=0,\ \ J^-_{i'j'} \epsilon_+= - {1 \over 2} \gamma_{i'j'} \epsilon_+ ;\ \ 
J^+_{i'j'} \epsilon_- = - {1 \over 2} \gamma_{i'j'},\ \  \epsilon_- \ \ J^+_{i'j'} \epsilon_-=0
 \eqno(5.5)$$
If we take  $\epsilon_+$ and  $\epsilon_-$ to be the $8_c$ and $8_s$ representations of ${\rm SO}(8)$ respectively then the spinor $\epsilon=\epsilon_+\oplus \epsilon_-$ is in the 
$(I\otimes 8_c)\oplus(8_s\otimes I) $ of ${\rm SO}(8) \times {\rm SO}(8)$. 
\par
We now wish to decompose the gravitino into representations of ${\rm SO}(8) \times {\rm SO}(8)$. The action of the SO(16) generators can be read off from equation (5.3) by taking the indices to take the values $i'.j'=3, \ldots 10$. The gravitino obeys the condition 
$\gamma^i \psi_i = \gamma \cdot \psi + \gamma_9 \psi_9 = 0 $ and so we can eliminate the $\psi_9$ component in terms of $\gamma\cdot \psi\equiv 
\gamma^{i'}\psi_{i'}$. 
We find that the  ${\rm SO}(8) \times {\rm SO}(8)$ generators 
$J_{i_1' i_2'}^\pm$ act on the gravitino $\psi_{k'}$ as 
$$
J^{\pm}_{i_1' i_2'} \psi_{j'} = - {1 \over 2} \gamma_{i_1' i_2'}(1 \mp \gamma_9) \psi_{j'} - 2 \eta_{j'[i_1'} \psi_{i_2']} \pm {1 \over 3} \gamma_9 \gamma_{i_1' i_2'} \gamma_{j'} \gamma \cdot \psi -  \gamma_9 \gamma_{i_1' i_2'} \psi_{j'} $$
$$
\mp {2 \over 3} \gamma_9 \eta_{j'[i_1'} \gamma_{i_2']} \gamma \cdot \psi \mp{2 \over 3} \gamma_9 \gamma_{j'[i_1'} \psi_{i_2']} \pm {4 \over 3} \gamma_9 \eta_{j' [i_1'} \psi_{i_2']} 
\eqno(5.6)$$
In deriving this equation we have used the identity 
$$
\gamma_{k_1\ldots k_n} = {(-1)^{{n(n+1)\over 2}} \over m!} \epsilon_{k_1\ldots k_n} {}^{j_1\ldots j_m} \gamma_9\gamma_{j_1\ldots j_m}
\eqno(5.7)$$
where $n+m=8$. 
While the action of $J^\pm_{i'j'}$ on $\gamma \cdot \psi$  follows from equation (5.6) and it is given by 
$$
J_{i_1' i_2'}^\pm \gamma \cdot \psi = - {1 \over 2} \gamma_{i_1' i_2'} \gamma \cdot \psi \mp {1 \over 6} \gamma_9 \gamma_{i_1' i_2'}  \gamma \cdot \psi \pm {4 \over 3} \gamma_9 \gamma_{[i_1'} \psi_{i_2']}. 
\eqno(5.8)  $$
\par
We now take the combination 
$$
\Psi_i =  \psi_i - {1 \over 2} \gamma_i \gamma \cdot \psi 
\eqno(5.9)$$
which transforms as 
$$
J_{i_1' i_2'}^\pm \Psi_k = - {1 \over 2} \gamma_{i_1' i_2'} (1 \pm \gamma_9) \Psi_{k'} - 2 \eta_{k' [ i_1'} (1 \mp \gamma_9) \Psi_{i_2']} 
\eqno(5.10) $$
\par
If we define $\Psi_k^\pm= {1\over 2} (1\pm \gamma_9)\Psi_k$ then equation (5.10) can be rewritten as 
$$
J_{i_1' i_2'}^+ \Psi^+_k=  - {1 \over 2} \gamma_{i_1' i_2'}  \Psi^+_k  ,\ \ J_{i_1' i_2'}^-\Psi^+_k = -2\eta_{k [i_1} \Psi^+_{i_2]} ,\ \ 
$$
$$
J_{i_1' i_2'}^+\Psi^-_k = -2\eta_{k [i_1} \Psi^-_{i_2]}  ,\ \ J_{i_1' i_2'}^- \Psi^-_k=  - {1 \over 2} \gamma_{i_1' i_2'}  \Psi^-_k 
\eqno(5.11) $$
As a result we find that $\Psi_k= \Psi_k^+ + \Psi_k^-$ belongs to the $ (8_c,8_v)\oplus (8_v,8_s)$ representation of ${\rm SO}(8) \times {\rm SO}(8)$. 
Thus both the fermionic and bosonic degrees of freedom belong to spinor representations of SO(16) and they have a rather similar decompositions into representations of ${\rm SO}(8) \times {\rm SO}(8)$.  Indeed the bosonic degrees of freedom belong to the $(8_v, 8_v) \oplus (8_c ,  8_s) $ and one can interchange the bosons and fermions by interchanging $8_v$ with $8_c$ for the first ${\rm SO}(8)$ factor. 
\par
It was observed in reference [18] that the fermionic degrees of freedom encoded in the gravitino carried a highly unfaithful representation of 
the Cartan involution algebra of the relevant algebra and as a result the representation should be annihilated certain generators. This paper also pointed out that this set of generators would form a subalgebra that was an ideal. We will now compute the lowest level   generators in the context of irreducible representation, that is $I_c(E_{9})$. Using equation (5.3) with the indices taking the values $3,\ldots, ,10$, we find, taking 
$\lambda=-1$  that 
$$
(S_{i_1i_2i_3} -{2\over 6!} \epsilon^{i_1i_2i_3} {}^{j_1\ldots j_6} S_{j_1\ldots j_6})\psi_k= 0
\eqno(5.12)$$
and that 
$$
(J_{i j} + {2 \over 8!} \varepsilon_{[i|}{}^{r_1 \ldots r_8} S_{r_1 \ldots r_8,|j]}) \psi_{k} =0
\eqno(5.13) $$
  These are precisely the same generators that annihilated the bosonic states. There are no doubt an infinite  similar equations at higher levels. 
 One could also take the choice $\lambda=1$ and find the same ideal if one systematically changed our definition  of $\gamma_9$ by introducing a minus sign. 
Thus seen from the perspective of the irreducible representations of  $I_c(E_{9})$, the bosonic and fermionic states have a remarkably similar structure. They  both belong to 128-dimensional spinor representations of SO(16), they have the same ideal which annihilates the two representations which carry a representation of  the coset of $I_c(E_{9})$ with the ideal $I$ whose coset ${I_c(E_{9})\over I}$  which is just  SO(16). 
\par
However, there is an important difference in the way the bosons and fermions appear. The original irreducible representation for the bosons contains an infinite number of fields whose number is reduced by duality relations, while for the fermions we just have the gravitino. This latter representation was not deduced from some general theory but by hand following the example of the maximal supergravity theories in lower dimensions.  It would seem natural to introduce higher level spinors and have these linked to the gravitino by duality relations. One possibility is to introduce the fields 
$$
\psi_\alpha,\ \psi_{i_1\ldots i_8 \alpha},\  \psi_{i_1\ldots i_9 \alpha},\ \psi_{i_1\ldots i_9, j_1\ldots j_8 \alpha},\ \ldots 
\eqno(5.14)$$
and corresponding duality relations 
$$
\psi_{i_1\ldots i_8 }= \gamma_{i_1\ldots i_8 j}\psi_j, \ldots 
\eqno(5.15)$$
It might be straightforward to extend this representation to $I_c(E_{11})$ and even $E_{11}$. This latter step could be possible if one exploited the fact that the algebras SL(D) do admit spinor representation. These have not been popular as they infinite dimensional, but this is what we need in this context. [25]. 
\medskip
{{\bf 6. The covariant formulation}}
\medskip
In  the above section we examined how the bosonic degrees of freedom of supergravity arose as the massless irreducible representation of 
$I_c(E_{11})\otimes_s l_1$. In particular we saw that they occurred as an irreducible  representation of the little group $I_c(E_{9})$ as carried by the fields of equation (2.9).  To find the full representation of $I_c(E_{11})\otimes_s l_1$ one has to carry out a boost. This procedure was discussed in reference [7]. In carrying out this last step one does not introduce any additional fields and so even though all the symmetries are present  they are not manifest. However, in field theory one usually requires representations that have some of the symmetries realised in a covariant manner. For example, for  the irreducible representations of the Poincare algebra one usually requires the Lorentz algebra to be manifest.  In our case we want the $I_c(E_{11})$ symmetry to be essentially manifest. How to achieve this  was very briefly outlined in reference [7] and in this section we will discuss this procedure  in detail.  We will also examine how  the duality relations and annihilation generators  which are present in the irreducible representation are inherited into the covariant formulation. 
\par
Before we begin we will briefly recall the general theory for obtaining  a covariant formulation  of an irreducible representation of $I_c(E_{11})\otimes_s l_1$. We will use  the same notation as in section four of reference [7]. Let the little algebra be denoted by ${\cal H}$ and then any group element $g$ of $I_c(E_{11})\otimes_s l_1$ can be written as $g= e^{\varphi\cdot S} h$ for $h\in {\cal H}$. Before the boost we choose a linear representation $u_\alpha(0)$ of ${\cal H}$ which we can take  to transform as 
 $$
 U(g) u_\alpha(0)= D(g^{-1})_\alpha{}^\beta u_\beta (0), \ \ L_A u_\alpha(0)= l^{(0)}_A u_\alpha(0)
  \eqno(6.1)$$
This representation will  contain the irreducible representation of the little algebra we are considering. To be more precise,   our irreducible representation of the little algebra is embedded in the representation $u_\alpha(0)$ which, as a result, must be a  reducible representation of ${\cal H}$. 
\par
The next step is to boost  the  representation  $u_\alpha(0)$ to find a representation of $I_c(E_{11})\otimes_s l_1$  in much  the same way as we boosted the irreducible representation of ${\cal H}$ (see section four of reference [7]). In particular, we take 
 $u_\alpha(\varphi)\equiv U(e^{\varphi\cdot S} ) u_\alpha (0)$ where the symbol $U$ denotes the action of the group element. However, $u_\alpha(\varphi)$ does not transform covariantly and so instead we consider the objects 
 $$
 {\cal A}_\alpha (\varphi) = D(e^{\varphi\cdot S}) _\alpha {}^\beta u_\beta(\varphi)= D(e^{\varphi\cdot S}) _\alpha {}^\beta
 U(e^{\varphi\cdot S} ) u_\beta (0)
 \eqno(6.2)$$
  It is straight forward to show that  $U_\alpha (\varphi) $  transforms covariantly,  and in particular  
  $$
  U(g)  {\cal A}_\alpha (\varphi) = D(g^{-1})_\alpha{}^\beta  {\cal A}_\beta  (\varphi') 
  \eqno(6.3)$$
  where $g\in I_c(E_{11})$ and  we have used the relation $g e^{\varphi\cdot S} = e^{\varphi' \cdot S} h(g, \varphi)$ with  $h(g, \varphi)\in {\cal H}$. 
  Although we have embedded the original irreducible representation into a bigger representation of ${\cal H}$ we still have to ensure that we still have the irreducible representation and as a result we have to impose projection conditions and,  for the massless case, equivalence relations. How to do this in general has yet to be understood but we will do it for the massless irreducible representation. 
 \par
 For comparison we very briefly give an account of the spin one massless  irreducible representation of the Poincare algebra.  We can choose all momenta to vanish except for $p^+= p_-=\sqrt{2}m$. The little algebra is  SO(D-2) since we have to take the generator $J_{+i}$ to vanish in order to have a unitary representation. The representation of SO(D-2)  is carried by the fields $A_i(0)$, $i = 1,\ldots , D-2$.  We embed this irreducible representation  into the SO(D-2) reducible representation $A_\mu (0),\ \mu=0,1\ldots ,D-1$. To get the same fields as in the original irreducible representation we demand that   $A_{+}(0) = 0$ and require  the equivalence relation   $A_{-}(0)\sim A_{-}(0)+ p_{-}\Lambda(0)$. The latter  allows us to set $A_{-}(0)=0$. To find the effect of these conditions  in the covariant formulation in terms of  $A_\mu$ we can impose these conditions in the rest frame fields, as given in equation (6.2),  and we recover, in $x$-space the familiar gauge fixing condition   $\partial^a A_a(x) = 0$ and the well known  gauge transformation   $A_a(x) \sim A_a(x) + \partial_a \Lambda(x)$. 
\par
We will now carry out the above discussion in the context of the massless irreducible representation we studied in the earlier sections. The first step is to embed the fields of  the irreducible of  of $I_c(E_{9}) \otimes_s l_1$ contained in equation (2.9)  into a larger representation. There is an obvious candidate, namely, we take the fields corresponding to the Cartan involution odd generators of $E_{11}$ rather than  just those of $E_{9}$. As such we consider the fields 
$$
{\cal A}_{\alpha} (0) = \{ h_{a_1 a_2}(0)  \ , \ A_{a_1 a_2 a_3}(0) \ , \ A_{a_1 \ldots a_6} (0)\ , \ h_{a_1 \ldots a_8,b} (0)\ , \ \ldots \} \ \ a,b,\ldots = 0,\ldots,10. 
\eqno(6.4) $$
which carry   a linear reducible  representation of $I_c(E_{11})$. These correspond to  the fields $ {\cal A}_\alpha (0) $ of equation (6.2). 
\par
We will now address the issue of the embedding condition that will ensure that we are really still dealing with the same massless irreducible representation.  We will first do this before the fields are boosted as in equation (6.2). We take  the  fields ${\cal A}_{\alpha}(0) $ of equation (6.4) to be subject to two conditions. The first of these is 
$$
K^{AB} G_{A,B}{}^C = K^{AB} (D^{\underline{\alpha}})_B{}^C \partial_A  {\cal A}_{\underline{\alpha}} = 0
 \eqno(6.5)$$
 where $K^{AB}$ is the metric  on a tangent space  which transforms under  $I_c(E_{11})$ as its indices suggest [13], and  the second is the equivalence relation 
 $$
 {\cal A}_{\underline{\alpha}} \simeq  {\cal A}_{\underline{\alpha}} + (D_{\underline{\alpha}} + D_{-\underline{\alpha}})_A{}^B \partial_B \Lambda^A. 
\eqno(6.6)$$
The matrices  $(D^{\underline{\alpha}})_B{}^C$ are those of the first fundamental representation and are defined in the equation  $[R^{\underline{\alpha}},l_B] = - (D^{\underline{\alpha}})_B{}^C l_C$.  As we will see equation (6.5)   eliminates all fields with  a lower $+$ index, while the second  equation (6.6) eliminating fields which take on a lower $-$ index, leaving us with the fields of equation (2.9), that is,  those of the original massless irreducible representation. 
\par
When acting on the fields of equation (6.4), the only component of $\partial_A$ which is non-zero is $\partial_-= {\partial \over \partial x^-}$. Thus the only non-zero component of $K^{AB}$ in equation (6.5)  is $K^{-+} = \eta^{-+}$, and so  equation (6.5) reduces to $(D^{{\alpha}})_{+}{}^C \partial_{-}  {\cal A}_{{\alpha}} = 0$. Since $p_-= \sqrt {2}m $ is a constant we have the condition $(D^{{\alpha}})_{+}{}^C  {\cal A}_{{\alpha}} = 0$. 
The above matrix occurs in the equation $[R^{{\alpha}},P_+] = - (D^{{\alpha}})_+{}^B l_B$  and so our condition can be expressed as $ [  {\cal A}_{\alpha} R^{\alpha} , P_+] =0$ which consists of transformations $ {\cal A}_{\alpha} \ R^{\alpha} $ that preserve the $p_+=0$ with the only non-zero momentum being $p_-$. By  definition this includes the little algebra transformations $E_{9}$ and so the condition of equation (6.5) places no constraints on the fields in equation (2.9). At levels zero, one and two    the condition of equation (6.5) lead to  
$$ 
h_{++}=0= h_{+i}= h^i{}_i; \ \ A_{+a_1a_2}=0= A_{+a_1\ldots a_5}
\eqno(6.7)$$
At higher levels we find that the commutator  $ [ A_{\alpha} R^{\alpha} , P_+] =0$  contains a factor $\delta^a_+$ times $ {\cal A}_{\alpha}$  times the  $l_1$ generator at the corresponding level. Hence  we find that the condition of equation (6.5) implies that any field $ {\cal A}_\alpha$ with a lower $+$ index is set to zero. 
\par
We can recover the same results by directly examining the component equations that  following from equation (6.5). At levels zero, one and three, these  are given by  [13]
$$
\partial^e h_e{}^a - {1 \over 2} \partial^a h_e{}^e = 0 \ \ \ , \ \ \ \partial^e A_{e a_1 a_2} = 0 \ \ \ , \ \ \ \partial^e A_{e a_1 \ldots a_5} = 0 \ \ , \ \ \ \ldots \eqno(6.8)$$
where we have thrown away derivatives with respect to the higher level coordinates.   By taking all the momentum except $p_-$ non-zero in equation (6.8) we  recover the above  result. 
\par
We now consider the equivalence relation   (6.6). In the rest frame, that is, for the fields of equation (6.4),   this  reduces to $(D_{{\alpha}} + D_{-{\alpha}})_A{}^- \partial_- \Lambda^A$ which is proportional to $ (D_{{\alpha}} + D_{-{\alpha}})_A{}^-\Lambda^A$. This matrix occurs in the commutator $[R^{{\alpha}} +R_{{\alpha}},l_A] = - (D^{{\alpha}}+D_{{\alpha}})_A{}^- P_-$. At level one we have the commutator 
$[R_{a_1 a_2 a_3},Z^{b_1 b_2}] = 6 \delta^{b_1 b_2}_{[a_1 a_2} P_{a_3]}$ and so we require $R_{a_1 a_2 -}$ in order to get $P_-$. At higher levels one also  finds that this expression will only be non-zero if the index ${{\alpha}}$ contains a lower $-$ index. To see this we note that the level zero  $P_-$ on the right-hand side  arises  from  the commutator of a level $n$  $l_1$ generator and a level $-n$ level generator of $E_{11}$ with a   resulting structure constant that  contains a $\delta _-^a$ factor. Hence  all the fields ${\cal A}_\alpha$ with a lower $-$ index are subject to an equivalence relation and for a suitable choice of $\Lambda^A$ they can be set to zero. 
 \par
 The equivalence relation (6.6) have been worked out at  low levels [14] and they lead to the expected gauge transformations;  
$$
\delta h_{(ab)}  = \partial_a \xi_b+  \partial_b \xi_a \ \ , \ \ \delta A_{a_1 a_2 a_3} =  \partial_{[a_1} \Lambda_{a_2 a_3]} \ \ , \ \ \delta A_{a_1 \ldots a_6} = 2 \partial_{[a_1} \Lambda_{a_2 \ldots a_6]} \ \ , \ \ \ldots 
\eqno(6.9)$$ 
As the only non-zero momenta is $p_-$ we find that at levels zero, one and two we can set to zero $h_{-+}$, $A_{-a_1a_2}$ and $A_{-a_1\ldots a_5}, \ldots $, in agreement with the above discussion. 
\par
So far we only considered the effect of the conditions of equation (6.5) and (6.6) in the rest frame, that is, $p_-=\sqrt {2} m$ all other members of the vector representation being zero. To find the analogue of these conditions for the covariant theory we take the conditions to act on $u_{\alpha}(0)$   in equation (6.2) and then apply the boost and matrix multiplication that this  equation contains. Since equation (6.5) contains the $I_c(E_{11})$ covariant expression $K^{AB} (D^{{\alpha}})_B{}^C \partial_A $ it retains the same form under the boost and so we can take this equation to hold for the theory after the boost. Taking equation (6.6) to hold in the rest frame and carrying out the boost  and matrix multiplication,as in equation (6.2), we find that it takes the form as in equation (6.6) but with a  parameter $\Lambda^A(\varphi)= D(e^{-\varphi\cdot S})_C{}^A U(e^{\varphi\cdot S})\Lambda^C(0)$. We recognise this as a gauge transformation of reference [14]. 
 \par
Hence we have shown that the conditions of equations (6.5) and (6.6) do not affect the fields of equation (2.9). However for the fields of equation 
(6.4) we find that they set all fields with a lower  $+$ to zero and we can, using the equivalence relation remove all fields with a lower $-$ index. As a result equations (6.5) and (6.6) ensure that $A_{\alpha}$,  given in equation (6.4), contains the same fields as occurred in the original massless irreducible representation,  given in equation (2.9). Hence it is just the same massless irreducible representation of $I_c(E_{11})\otimes_s l_1$. 
\par
In section two we found that the fields of the massless irreducible representation obeyed the duality relations of equation (2.19), (2.20) and (2.22) as well as similar higher level equations. While  in section three  we found that as a result  this  representation was annihilated by the generators which belonged to an ideal, for example those in equations (3.6) and (3.11). As the covariant formulation of the irreducible representation is essentially  the same as the original  irreducible representation we can expect that the generators in equations  (3.6) and (3.11),  as well as those at  higher level, imply the presence of analogous equations which we will not find. Indeed, 
starting from the equations that  the  generators in the ideal $I$ annihilate the fields in the irreducible representation in the rest frame,  we can carry out a boost, as in equation (6.2),   to find an infinite number of   constraints in the covariant theory. However, as the boost contains  generators which levels greater than zero it will transform  the elements in the ideal with different levels into each other.  
\par
Rather than carry out the boost,  we will find some covariant operators that agree with elements in the ideal in the rest frame.  We begin by considering the operator 
$$
\hat N_{a_1 \ldots a_4} = P_{[a_1} S_{a_2 a_3 a_4]} + {1 \over 2.6!} \varepsilon_{a_1 \ldots a_4}{}^{c_1 \ldots c_7} P_{c_1} S_{c_2 \ldots c_7} + {1\over 4} Z^{e_1 e_2} S_{e_1 e_2 a_1 \ldots a_4} 
$$
$$
+{1\over 7.2!}\epsilon_{a_1 \ldots a_4}{}^ {c_1\ldots c_7}Z^{e_1e_2}S_{e_1c_1\ldots c_7, e_2} 
+{1\over 12} Z_{[a_1a_2} J_{a_3a_4]}
\eqno(6.10)$$
 Acting on the fields of the massless irreducible representation of equation (2.9), which are in the rest frame,  it obviously vanishes except for $\hat N_{-i_1i_2i_3}$ which is equal to $ p_{-}N_{i_1i_2i_3}$, where  $N_{i_1i_2i_3}$ is defined in equation (3.6). However, as we have shown in section three $N_{i_1i_2i_3}$   vanishes on the  
massless irreducible representation of equation (2.9) and as a result $\hat N_{a_1 \ldots a_4} $ also vanishes on this representation on the fields  the rest frame. The terms in $\hat N_{a_1 \ldots a_4} $ that contain higher level elements of $l_1$ vanish in the rest frame and the precise coefficients that we have given will be derived below. Clearly, $\hat N_{a_1 \ldots a_4} $ is one of the covariant operators resulting from the action of the  boost on the annihilation operators in the rest frame as discussed just above.
\par
We now consider if $\hat N_{a_1 \ldots a_4} $ really does annihilate the covariant states of equation (6.4).  Since the covariant fields are only defined up to an equivalence, or equivalently gauge symmetry, it simplifies the calculation considerably if we act on objects that are gauge invariant. At the linearised level such objects are 
$$
\omega_{a,b_1 b_2} = - \partial_{b_1} h_{(b_2 a)} + \partial_{b_2} h_{(b_1 a)} + \partial_a h_{[b_1 b_2]}, \ 
G_{a_1 \ldots a_4} = \partial_{[a_1} A_{a_2 a_3 a_4]},   \ G_{a_1 \ldots a_7} = \partial_{[a_1} A_{a_2 \ldots a_7]},  \ \ldots 
\eqno(6.11)$$
We find that 
$$
\hat{N}^{b_1 b_2 b_3 b_4} G_{a_1 a_2 a_3 a_4} = - {1 \over 4!}  P_{a_1} \varepsilon_{a_2 a_3 a_4}{}^{b_1 .. b_4}{}^{c_1 c_2 c_3 c_4} E_{c_1 c_2 c_3 c_4} +  3 \delta^{[b_1 b_2}_{a_1 a_2} P_{a_3} E_{a_4],}{}^{b_3 b_4]}  
\eqno(6.12) $$
and 
$$
\hat{N}^{b_1 b_2 b_3 b_4} \omega_{a_1 , a_2 a_3} = - 4 \eta_{a_1 a_3} P_{a_2} E^{b_1 \ldots b_4} + 4 \eta_{a_1 a_2} P_{a_3} E^{b_1 \ldots b_4} - 90 P_{c_1} \eta_{c_2 a_3} \delta^{[c_1 c_2}_{a_1 a_2} E^{b_1 \ldots b_4]}  $$
$$
- 90 P_{c_1}  \eta_{c_2 a_2} \delta^{[c_1 c_2}_{a_3 a_1} E^{b_1 \ldots b_4]} - 90 P_{c_1}  \eta_{c_2 a_1} \delta^{[c_1 c_2}_{a_3 a_2} E^{b_1 \ldots b_4]}  $$
$$
+ \partial_{a_1} \tilde{\Lambda}_{a_2 a_3}{}^{b_1 \ldots b_4}  
\eqno(6.13) $$
where 
$$
E_{a_1 \ldots a_4} = G_{a_1 \ldots a_4} - {1 \over 2 \cdot 4!} \varepsilon_{a_1 \ldots a_4}{}^{b_1 \ldots b_7} G_{b_1 \ldots b_7} = 0,
 \eqno(6.14) $$
$$E_{a,b_1 b_2} = \omega_{a,b_1 b_2} - {1 \over 4} \varepsilon_{b_1 b_2}{}^{c_1 \ldots c_9} G_{c_1 \ldots c_9,a} = 0, 
\eqno(6.15)$$
and 
$$
\partial_{a_1} \tilde{\Lambda}_{a_2 a_3}{}^{b_1 \ldots b_4} = P_{a_1} [ 12 \delta_{a_3}{}^{[b_1} G_{a_2}{}^{b_2 b_3 b_4]} - 12 \delta_{a_2}{}^{[b_1} G_{a_3}{}^{b_2 b_3 b_4]} + 3  \delta_{a_3}{}^{[b_1} P_{a_2} A^{b_2 b_3 b_4]} $$
$$
- 3 \delta_{a_2}{}^{[b_1} P_{a_3} A^{b_2 b_3 b_4]} + {3 \over 8} \varepsilon^{b_1 \ldots b_4}{}^{c_1 \ldots c_7} P_{c_1} \eta_{c_2 a_2} A_{a_3 c_3 \ldots c_7} 
- {3 \over 8} \varepsilon^{b_1 \ldots b_4}{}^{c_1 \ldots c_7} P_{c_1} \eta_{c_2 a_3} A_{a_2 c_3 \ldots c_7}] 
\eqno(6.16)$$
We recognise equations (6.14) and (6.15) as the duality relations that occur  in the covariant theory arsing from the non-linear realisation of 
$E_{11}\otimes _s l_1$ with local subalgebra $I_c(E_{11})$ [ 1,2,3,4]. Thus we find that $N_{a_1 \ldots a_4} $ will indeed vanish on  the covariant  fields  provided the duality relations hold. 
The last term in equation (6.13) is defined in equation (6.16) and it  just corresponds to the fact that $\omega_{a,b_1 b_2}$ is subject to local Lorentz transformations. We note that   $\tilde{\Lambda}_{a_2 a_3}{}^{b_1 \ldots b_4}$ is indeed of the form of a local Lorentz transformation since the $b_1 \ldots b_4$ indices can be contracted with a parameter with these indices. 
\par
We will now consider the operator 
$$
\hat{N}_{b_1 b_2 b_3} = P_{[b_1} J_{b_2 b_3]} - {1 \over 8!} \varepsilon_{[b_1 b_2|}{}^{e_1 \ldots e_9} P_{e_1} S_{e_2 \ldots e_9,|b_3]} +\ldots \eqno(6.17)$$
where $+\ldots$ means terms which contain higher order $l_1$ generators. 
Acting on the fields of equation (2.9) in the rest frame this operator we find that it  obviously vanishes except for 
$\hat{N}_{-i_2 i_3} = p_{-} N_{i_2i_3}$  where $N_{i_2i_3}$ is defined in equation (3.11). However, this generator also vanishes on the rest frame states and so  $\hat{N}_{b_1 b_2 b_3}$ vanishes on the rest frame states. As such we would expect it to vanish on the covariant fields of equation (6.4)
One finds that 
$$
\hat{N}^{b_1 b_2 b_3} \omega_{a_1,a_2 a_3} = \eta_{a_1 a_3} P_{a_2} E^{[b_1,b_2 b_3]} - P_{a_2} \delta^{[b_1}{}_{(a_3} E_{a_1),}{}^{b_2 b_3]} - \eta_{a_1 a_2} P_{a_3} E^{[b_1,b_2 b_3]} $$
$$
+ P_{a_3} \delta^{[b_1}{}_{(a_2} E_{a_1),}{}^{b_2 b_3]}$$
$$
- 6 P_{r_1} (\eta_{r_2 a_3} \delta^{[b_1 b_2}_{a_2 a_1} E^{|b_3|,r_1 r_2} + \eta_{r_2 a_1} \delta^{[b_1 b_2}_{a_2 a_3} E^{|b_3|,r_1 r_2]} - \eta_{r_2 a_2} \delta^{[b_1 b_2}_{a_3 a_1} E^{|b_3|,r_1 r_2]})$$
$$
+ \partial_{a_1} \tilde{\Lambda}_{a_2 a_3}{}^{b_1 b_2 b_3} = 0, 
\eqno(6.18)$$
where 
$$
\partial_{a_1} \tilde{\Lambda}_{a_2 a_3}{}^{b_1 b_2 b_3}  = P_{a_1} \{ \varepsilon^{[b_1 b_2|r_1 \ldots r_9} P_{r_1}( \eta_{r_2 a_3} h_{a_2 r_3 \ldots r_9}{}^{,b_3]} - \eta_{r_2 a_2} h_{a_3 r_3 \ldots r_9}{}^{,b_3]}) $$
$$
+ P^{[b_2} ( \delta^{b_1}{}_{a_2} h_{a_3}{}^{b_3]} - \delta^{b_1}{}_{a_3} h_{a_2}{}^{b_3]}) \}
 \eqno(6.19) $$
 is an  expected Lorentz transformation. 
\par
Since the operators $\hat N_{a_1 \ldots a_4}$ and $\hat N_{a_1 a_2 a_3}$, and presumably their higher level analogues, annihilate the covariant fields one would expect their commutator with the generators of $I_c(E_{11})$ to give them back. At lowest order, and assuming we can find a missing factor of two,  we find that 
$$
[S_{a_1 a_2 a_3},\hat{N}_{b_1 b_2 b_3 b_4}] 
= +9.3 \delta ^{[a_1a_2}_{[b_1b_2} \hat N^{a_3]}{}_{b_3b_4]}
 +{1\over 24} \epsilon _{b_1b_2b_3b_4}{}^{a_1a_2a_3}{}^{c_1c_2c_3c_4}\hat N_{c_1c_2c_3c_4}
\eqno(6.20) $$
Similar results can be expected for  $\hat N_{a_1 \ldots a_3}$ and  the higher level generators of $I_c(E_{11})$. It would be interesting to compute the commutators of $\hat N_{a_1 \ldots a_4}$ and $\hat N_{a_1 \ldots a_3},\ldots $ with themselves. Given the above result one would expect the result to be proportional to $\hat N$'s times $l_1$ generators. A result which would again annihilate the covariant fields. 
\par 
The covariant  fields are subject to the Casimir condition $L^2 \Psi = K^{AB} L_A L_B \Psi = 0$ as well as higher level conditions [15]. We would expect that the generators $\hat N_{a_1 \ldots a_4}$ and $\hat N_{a_1 \ldots a_3},\ldots $ will commute with these up to terms proportional to themselves. They should also be consistent with the gauge transformation of equation (6.5). 
\par
In this section we have found that the generators in the ideal $I$ that annihilate the massless irreducible states of equation (2.9) in the rest frame lead to an infinite set of operators which annihilate the massless  irreducible representation when expressed in terms covariant fields provided  the covariant fields obey  an infinite set of duality relations. These covariant duality relations  are precisely  those that occur in the non-linear realisation of  $I_c(E_{11})\otimes_s l_1$ with local subalgebra  $I_c(E_{11})$. This is consistent with the observation that the duality conditions on the covariant fields are just covariant versions of the duality identities that exist in the original massless irreducible representation [7]. Thus the  existence of the duality relations which contain the dynamics arise from the irreducible representation and in particular many of them correspond to the action of the affine action  of $I_c(E_9)$  on the  $I_c( E_8)$ representation of equation (2.24). 

\medskip
{{\bf 7. An alternative formulation of fermions in E theory }}
\medskip
The spinors were introduced   in E theory by hand in the sense that they do not follow in natural way from the $E_{11}$ algebra but started from  the familiar gravitino and insisted that it carry a representation of $I_c(E_{11})$, which was constructed by hand.  In this section we will take a different approach which generalises, in a natural way, the way spinors appear in the context of the Poincare algebra.  For  the Poincare algebra, that is,  $SO(1, D-1)\otimes_s T^D$ the vector representation $T^D$ leads to a $D$ dimensional spacetime on which $SO(1, D-1)$ acts. For each coordinate of the spacetime we introduce a $\gamma^a$ matrix which act on a spinor. 
\par
In E theory we start from the  algebra $I_c(E_{11})\otimes_s l_1$  which is also    a semi-direct product. The  vector representation $l_1$ leads in the nonlinear realisation to  the spacetime  and we now introduce matrices $\Gamma^A$'s which are in one to one correspondence with the vector  representation, hence their label $A$. As a result they are also carry the indices of the spacetime  coordinates in E theory.  For example, in eleven dimensions we introduce the matrices 
$$
\Gamma^A=\{ \Gamma^a,\ \Gamma^{a_1a_2} ,\ \Gamma^{a_1\ldots a_5}, \  \Gamma^{a_1\ldots a_7,b}, \ \Gamma^{a_1\ldots a_8}, \ldots \}
\eqno(7.1)$$
\par
These matrices are to obey the equation 
$$
\Gamma^A \Gamma^B+ \Gamma^B \Gamma^A= 2K^{AB}
\eqno(7.2)$$
where $K^{AB}$ is the $I_c(E_{11})$ invariant tangent space metric. We now introduce a spinor $\Psi$ which carries a representation of the $\Gamma$ matrices and transforms under $I_c(E_{11})$ as 
$$
\delta \Psi= U( S^\alpha ) \Psi= -{\cal S}^\alpha \Psi
\eqno(7.3)$$
where $U (S^\alpha )$ is the action of the generator $S^\alpha$ and the matrix ${\cal S}^\alpha$ its effect. We require that 
$$ 
[{\cal S}^\alpha , \Gamma^A] =  \Gamma^B (\tilde D^\alpha )_B{}^A{}
\eqno(7.4)$$
where  $\tilde D^\alpha\equiv  D^\alpha-  D_\alpha$. By its definition the matrix of the vector representation appears in the commutator  $ [R^\alpha , l_A ]=- ( D^\alpha )_A{}^B l_B$. We recall that  $S^\alpha= R^\alpha- R_\alpha$ and that $(\tilde D^\alpha )_A{}^CK_{CB}$ is an antisymmetric matrix. 
As such the commutator of  ${\cal S}^\alpha$ with   the $\Gamma^A$'s leads to a  transform  under $I_c(E_{11})$ which is that of  the vector representation. 
\par
A generalised Dirac equation is given by 
$$
\Gamma^A \partial_A \Psi= 0
\eqno(7.5)$$
It is invariant under $I_c(E_{11})$ transformations as the derivatives $\partial_A\equiv {\partial\over \partial x^A}$ transforms as $\delta (\partial_A)=-(\tilde D^\alpha )_A{}^B \partial_B$.  
\par
At level zero $I_c(E_{11})$ is just SO(1,10) and so at this level the above discussion just reduces to  the standard discussion of the Dirac equation. The Gamma matrices corresponding to the higher level coordinates occur with derivatives with respect to these coordinates. Thus if we neglect the higher level derivatives this is just the familiar Dirac equation. We note that we can not take $\Gamma^{a_1a_2}, \ldots $ to be proportional to  the standard gamma matrix $ \gamma^{a_1a_2} , \ldots $  as this does not satisfy equation (7.2)
\par
The above applies if we replace $E_{11}$ by any Kac-Moody algebra and the vector representation by anyone of its representations. To give a simple example we consider the non-linear realisation of $A_1^{+++}\otimes l_1$ which corresponds to gravity in four dimensions. At levels zero and one the  generators of the vector representation are $P_a$ and $Z^a$ and so we introduce the gamma matrices $\Gamma^A=\Gamma^a , \tilde \Gamma^a, \ldots $ with  $a=0,1,\ldots , 3$. The algebra 
$I_c (A_1^{+++})$ contains the generators $J_{ab}$ and $S_{ab}$ at levels zero and one respectively. For an account of this theory see reference [26] and the earlier references it contains. We can take the gamma matrices to be 
$$
\Gamma^a =\gamma^a\otimes I , \quad 
\tilde \Gamma^a =2i\gamma^5\otimes \gamma^a
\eqno(7.6)$$
where $\gamma^a\gamma^b+ \gamma^b\gamma^a=2\eta^{ab}$ are the usual gamma matrices in four dimensions, $(i\gamma^5)^2= I$ and $\gamma^5\gamma^b+ \gamma^b\gamma^5=0$. Thus this spinor has eight components. 
\par
The vector representation of the algebra $I_c(E_{11})$ contains at level one the coordinates $x^{a_1a_2}$ and so the corresponding spinor will carry a representation of the matrices $\Gamma^{a_1a_2}$ which obey 
$$
\Gamma^{a_1a_2}\Gamma_{b_1b_2} +\Gamma_{b_1b_2}\Gamma^{a_1a_2}= 2\delta _{b_1b_2}^{a_1a_2}
\eqno(7.7)$$
\par
Such matrices are often required if we consider other very extended Kac-Moody algebras. If we have the index range  in three dimensions then we can define  $\gamma^a= {1\over 2} \epsilon^{a b_1b_2}\Gamma_{b_1b_2}$ and we can use our usual representation of gamma matrices in three dimensions. So we require the spinor in this sector to have two components. If the index range is in four, for example,  Euclidean dimensions then we can define 
$$
\Gamma^{+} {}^{a_1a_2}= {1\over 2}(\Gamma^{+} {}^{a_1a_2}\pm {1\over 2}  \epsilon ^{a_1a_2b_1b_2}\Gamma_{b_1b_2})
\eqno(7.8)$$
which obey the relations 
$$
 {1\over 2} \epsilon ^{a_1a_2b_1b_2}\Gamma^{\pm}_{b_1b_2}= \pm \Gamma^{\pm} {}^{a_1a_2},\ \ {\rm and} \ \ 
  \{\Gamma^{+} {}^{a_1a_2}, \Gamma^{-} {}_{b_1b_2}\}= \delta _{b_1b_2}^{a_1a_2}
 \eqno(7.9)$$
 As such  we can take the independent matrices  to be $\Gamma^{\pm} {}^{12}$, $\Gamma^{\pm} {}^{13}$ and $\Gamma^{\pm} {}^{14}$ and a representation is formed by defining the vacuum $|>$ to satisfy 
 $$
 \Gamma^{-} {}_{12} |>=0= \Gamma^{-} {}_{13} |>= \Gamma^{-} {}_{14}|>
 \eqno(7.10)$$ 
 We can regard the matrices $\Gamma^{+} {}^{12}$, $\Gamma^{+} {}^{13}$ and $\Gamma^{+} {}^{14}$ as creation operators acting on this vacuum, that is, 
 $$
 |>, \  \Gamma^{+} {}_{12} |>, \ldots ,  \Gamma^{+} {}_{12}  \Gamma^{+} {}_{13}|>,  \ \ldots 
 \eqno(7.11)$$ 
Thus we have $(1+1)^3=8$ states, or spinor components, due to this  sector of the spinor. 
\par
Let us also consider the algebra $I_c(E_{8})\otimes_s l_1$. The vector representation of $E_{8}$ has dimension 248 and as explained in section 
four this decomposes into the 120 plus 128 dimensional representations of $I_c(E_{8})={\rm SO}(16)$. We can take Gamma matrices corresponding to both of these representations or just one of them. If we take just the spinor 128 dimensional representation the we should take the Gamma matrices 
$$
\Gamma^i , \ \Gamma^{i_1i_2} , \ \Gamma^{i_1i_2i_3}  , \ \Gamma^{(ij)} , \ \ {\rm where}\ \ i,j=1,2,\ldots , 8
\eqno(7.12)$$
where we have given them in the ${\rm SO}(8)\otimes {\rm SO} (8)$ decomposition of section four. The generalised Dirac equation would have the form 
$$
(\Gamma^i \partial_i + \Gamma^{i_1i_2} \partial_{i_1i_2} +\ldots )\Psi=0
\eqno(7.13)$$
The Gamma matrices can be written in block diagonal form using the analogue of the $\gamma_5$ matrices as above. The  $\Gamma^i$ can be taken to be the usually gamma matrices in the first block and as their are eight of them this part of the spinor will have $2^4=16$ components. More generally the spinor which carries a representation can be found by defining annihilation and creation operators. We might expect 64 creation operators and 64 destruction operators and so the corresponding part of the  spinor should have $2^{64}$ components. 
\par
The above considerations were at the linearised order but it is straightforward to generalise it to the full symmetries of the $E_{11)}\otimes_s l_1$  by taking 
$$
\Gamma^A E_A{}^M (\partial_M + Q_M)\Psi=0
\eqno(7.13)$$
where $E_M{}^A$ is the vielbein and $Q_M$ the connection found in the non-linear realisation. 
\par
We end this section with some speculative remarks. To account for the  gravitino we can introduce  the object $\Psi_A$ which is a spinor with the  vector index $A$. We could take this to obey the on-shell conditions 
$$
\Gamma^B \partial_B \Psi_A=0= \Gamma^A\Psi_A= \partial^A\Psi_A
\eqno(7.14)$$
which does indeed contain the correct on-shell states if we restrict to the first part of the spinor $\Psi_a$. We leave it to the future to examine how this fits into the full theory. 
\par
We could also  introduce a generalised supersymmetry generator $Q^A$ which is a generalised  spinor  and  so with  $I_c(E_{11})$ generators it has the relation
$$
[S^\alpha , Q]= -{\cal S}^\alpha Q
\eqno(7.6)$$
One might speculate that they obey an  anti-commutator of the generic form
$$
\{Q , Q\}= \Gamma^A l_A
\eqno(7.7)$$
\par
How the generalised spinor introduced in this section fits in detail into E theory and what are its mathematical properties for  future study. 
\medskip
{{\bf 8 Discussion}}
\medskip
In this paper we have analysed in detail the irreducible representation of $I_c(E_{11})\otimes_s l_1$ corresponding to a massless point particle.  The corresponding  little algebra, $I_c(E_{9})\otimes_sl_1$,  is an infinite dimensional Lie algebra and  as a result one may expect it to contain an  infinite number of degrees of freedom. However, the fields  are subject to an infinite number of duality relations which are preserved by $I_c(E_{9})$ and these reduce the number of independent degrees of freedom to be just  128  which  belong to the spinor representation of  $I_c(E_{8})={\rm SO}(16)$. These can    be taken to be $h_{i j } \ (h^i{}_i=0$) and $A_{i_1i_2i_3}$ where $i,j=2,\ldots , 10$.  They are indeed the  bosonic degrees of freedom of eleven dimensional supergravity. The infinite number of duality equations relate all the other fields in the representation to these fields.
A consequence of these duality relations is that the irreducible representation is annihilated by an infinite number  of generators of  $I_c(E_{9})$ which form a subalgebra $I$ which is an ideal. The Lie algebra 
$I_c(E_{9})\over I$ is SO(16). 
\par
The 128 bosonic  independent degrees of freedom  belong to the $(8_v, 8_v)\oplus (8_c ,  8_s)$ representations when decomposed into the subalgebra  ${\rm SO}(8) \times {\rm SO}(8)$ subalgebra of SO(16). These two representations contain the fields ($h_{i'j'}$, $A_{i'_1\ldots i'_6} $) and  ($A_{i'_1i'_2i'_3}$,  $h_{i'_1 \ldots i'_8, k'}$)   respectively where $i',j'=3,\ldots ,10$. The SO(16)  transformations not in the  ${\rm SO}(8) \times {\rm SO}(8)$ subalgebra change these representations into each other. The remaining fields which appear in the irreducible representation of $I_c(E_9)$ are  affine  copies of these 128  fields and are related by duality relations to the fields in the $(8_v, 8_v)\oplus (8_c ,  8_s)$ representations. 
Like all such irreducible representations we can formulate the massless irreducible representation in a covariant manner. The infinite number of duality relations become covariant duality equations which contain the dynamics and one finds a corresponding set  of operators that annihilate the representation. 
\par
The dynamics that results from the non-linear realisation of $E_{11}\otimes_s l_1$ with local subalgebra $I_c(E_{11})$ has been found at low levels and the resulting equations of motion agree precisely with those  of maximal supergravity if one discards the dependence on the spacetime coordinates beyond those usually considered. The dynamics appears through an infinite set of duality equations that relate an infinite number of the higher level fields to the graviton and three form and it is by taking space time derivatives of  these that one can find the standard equations of motion of maximal supergravity.  The duality relations that arise in the covariant formulation of the massless irreducible of  $I_c(E_{11})\otimes_s l_1$ in the covariant formulation  are essentially  those that arise in the non-linear realisation of $E_{11}\otimes_s l_1$ with local subalgebra $I_c(E_{11})$ at the linearised level. This provides strong support for the strongly suspected fact that the only degrees of freedom contained in the non-linear realisation  are the 128 bosonic degrees of freedom of supergravity. We also see that  the  infinite number of duality relations that appear in the non-linear realisation are a consequence of the duality relations that occur in the irreducible representation and this allows us to predict the structure of the duality relations in the former theory.  In the non-linear realisation there are  also higher level fields which have blocks of ten, or eleven,  indices and so these fields do not appear in the irreducible representation. These fields obey equations  predicted by the  non-linear realisation and they  lead to the gauged supergravities.
\par
In the irreducible representations in the E theory approach the bosonic and fermionic degrees of freedom appear in a unified way. They belong to the two different spinor representations of $I_c(E_{8})={\rm SO}(16)$ which are associated with the two nodes at the end of the SO(16) Dynkin diagram. Thus swopping 
thee two nodes results in swopping bosons and fermions. One puzzle with E theory is the way it leads to predictions that were usually seen as a result of supersymmetry. Examples are the appearance of the two and five brane charges and the BPS conditions [15]. 
Supersymmetry was discovered, at least from the Russian viewpoint, by demanding that internal and spacetime symmetries were contained in the same symmetry algebra. This required the introduction of the supersymmetry generators. However, $E_{11}$ achieves the same objective as it does contain the symmetries of spacetime, such as Lorentz symmetry, and also internal symmetries such as the internal exceptional symmetries of the maximal supergravities. 
\par
In  this paper we have concentrated on the massless irreducible representation of $E_{11}\otimes_s l_1$  which corresponds to the maximal supergravity theories. The vector representation of $E_{11}$ contains the brane charges and by taking these to be non-zero we will find other  irreducible representations of $E_{11}\otimes_s l_1$ corresponding to branes. In a future paper we hope to examine some of  these representations in detail and make the connection to the the brane dynamics as also appears in the context of E theory, see for example references [27].  
\par
The  interesting paper [28]   also remarks on the similar properties of bosons and fermions although from a  different viewpoint. It noticed that the SO(9) representations to which bosonic ($44\oplus 84$)and fermionic (128)  degrees of freedom emerge in a very natural way as a  solution of the equation $K\Psi=0$. In this equation  $K$ is the Kostant operator which is of the form $K=\sum_a T_a\gamma^a$ where $\gamma^a$ are gamma matrices and $T^a$ are generators of $F_4$ that belong to the coset ${F_4\over {\rm SO}(9)}$. The algebra $F_4$, dimension 52,  can be found from the algebra SO(9), dimension 36,   by adding to the latter the 16 generators that belong to the spinor representation of SO(16). 
\par
One can speculate that this picture can be generalised to incorporate the features of the irreducible representations found in this paper. 
The bosonic degrees of freedom belong  to one spinor representation of SO(16),  while the fermionic degrees of freedom belong to the other spinor representation,  both of which have dimension 128. The algebra $E_8$ emerges from the SO(16) algebra if we add to the 128 generators belonging to the spinor representation of SO(16) to the 120 generators of SO(16). We can then consider the coset of $E_8\over {\rm SO}(16)$. The corresponding Kostant operator would consist of   128 Gamma matrices multiplied by the generators in the coset. Could it be that the solutions of the equation $K\Psi=0$ contain the two 128 dimensional spinor representation of SO(16)? We can also wonder what are the higher spin solutions? Speculating even further we can suppose that some of these 128 Gamma matrices can be taken to be the supercharges similar to what took place in the discussion of the $N=2$ hypermultiplet given in reference [28].  We hope to study these matter further and how they might fit into E theory. 
\medskip
{{\bf Acknowledgements}}
\medskip
Peter West wishes to thank the SFTC for support from Consolidated grants number ST/J002798/1 and ST/P000258/1, while Keith Glennon would like to thank Kings College  for support during his PhD studies. 
\medskip
{{\bf References}}
\medskip
\item{[1]} P. West, {\it $E_{11}$ and M Theory}, Class. Quant.  
Grav.  {\bf 18}, (2001) 4443, hep-th/ 0104081. 
\item{[2]} P. West, {\it $E_{11}$, SL(32) and Central Charges},
Phys. Lett. {\bf B 575} (2003) 333-342,  hep-th/0307098. 
\item {[3]} A. Tumanov and P. West, {\it E11 must be a symmetry of strings and branes},  arXiv:1512.01644.
\item{[4]} A. Tumanov and P. West, {\it E11 in 11D}, Phys.Lett. B758 (2016) 278, arXiv:1601.03974.
\item{[5]} P. West, {\it Introduction to Strings and Branes}, Cambridge University Press, 2012.
\item{[6]} P. West,{\it  A brief review of E theory}, Proceedings of Abdus Salam's 90th  Birthday meeting, 25-28 January 2016, NTU, Singapore, Editors L. Brink, M. Duff and K. Phua, World Scientific Publishing and IJMPA, {\bf Vol 31}, No 26 (2016) 1630043, arXiv:1609.06863. 
\item{[7]} P. West,  {\it  Irreducible representations of E theory},  Int.J.Mod.Phys. A34 (2019) no.24, 1950133,  arXiv:1905.07324. 
\item{[8]}  E. P. Wigner, ÒOn unitary representations of the inhomogeneous Lorentz group,Ó Annals Math.40(1939) 149.
\item{[9]}  A. Kleinschmidt and P. West, {\it  Representations of G+++
and the role of space-time},  JHEP 0402 (2004) 033,  hep-th/0312247.
\item{[10]} P. West,  {\it $E_{11}$ origin of Brane charges and U-duality
multiplets}, JHEP 0408 (2004) 052, hep-th/0406150. 
\item{[11]} P. Cook and P. West, {\it Charge multiplets and masses
for E(11)}, ÊJHEP {\bf 11} (2008) 091, arXiv:0805.4451.
\item {[12]} P. West, {\it The IIA, IIB and eleven dimensional theories 
and their common
$E_{11}$ origin}, Nucl. Phys. B693 (2004) 76-102, hep-th/0402140. 
\item{[13]} M. Pettit and Peter West, {\it An E11 invariant gauge fixing}, Int.J.Mod.Phys. A33 (2018) no.01, 1850009, arXiv:1710.11024. 
\item{[14]} P. West, {\it Generalised Space-time and Gauge Transformations }, JHEP 1408 (2014) 050, arXiv:1403.6395. 
\item{[15]} P. West, {\it Generalised BPS conditions}, Mod.Phys.Lett. A27 (2012) 1250202,  arXiv:1208.3397. 
\item{[16]} E. Cremmer and B. Julia, {\it The SO(8) supergravity}, Nucl. Phys.B 159(1979) 141
\item{[17]} S. de Buyl, M. Henneaux and  L. Paulot, {\sl Extended E8
Invariance of 11-Dimensional Supergravity} JHEP {\bf 0602} (2006) 056
{\tt hep-th/05122992}
\item{[18]} T. Damour, A. Kleinschmidt qand  H. Nicolai {\sl
Hidden symmetries and the fermionic sector of eleven-dimensional
supergravity} Phys. Lett. B {\bf 634} (2006) 319 {\tt hep-th/0512163}.
\item{[19]} S. de Buyl, M. Henneaux and  L. Paulot {\sl Hidden
Symmetries and Dirac Fermions}Ê Class. Quant. Grav. {\bf 22} (2005) 3595
{\tt hep-th/0506009}.
\item{[20]} M Henneaux, E Jamsin, A Kleinschmidt and  D Persson; Phys.
Rev. D (2009) 045008; arXiv:0811.4358
\item{[21]}  D. Steele and P. West, {\it E11 and Supersymmetry}, JHEP 1102 (2011) 101, arXiv:1011.5820.
\item{[22]} A. Kleinschmidt, H. Nicolai, and J. Palmkvist, {\it $K(E_9)$ from $K(E_{10})$},Ó JHEP 06 (2007) 051, hep-th/0611314
\item{[23]} A. Kleinschmidt, {\it Unifying R-symmetry in M-theory}, in V. Sidoravi{\v c}ius (ed.) New Trends in Mathematical Physics, Proceedings of the XVth International Congress on Mathematical Physics, Springer (2009). hep-th/0703262.
\item{[24]} A. Kleinschmidt, H. Nicolai, and A.  Vigan{\`o}. {\it On Spinorial Representations of Involutory Subalgebras of KacÐMoody Algebras}, in Partition Functions and Automorphic Forms, pp. 179-215. Springer, Cham, (2020). arXiv:1811.11659
\item{[25]} Y. Ne'eman, {\it Gravitational interaction of hadrons Band-spinor representations of GL(n,R)}, Proc.Natl.Acad.Sci.USAVol.74,No.10,pp.4157-4159. 
\item{[26]} K. Glennon and P. West,  {\it Gravity, Dual Gravity and A1+++},  Int.J.Mod.Phys.A 35 (2020) 14, 2050068, arXiv:2004.03363. 
\item{[27]} P.  West,  {\it E11, Brane Dynamics and Duality Symmetries}, Int.J.Mod.Phys. A33 (2018) no.13, 1850080, arXiv:1801.00669; {\it A sketch of brane dynamics in seven and eight dimension using E theory}, Int.J.Mod.Phys. A33 (2018) no.32, 1850187, 
arXiv:1807.04176. 
\item{[28]} P. Ramond, {\it Boson-Fermion Confusion: The String Path To Supersymmetry},  Nucl.Phys.Proc.Suppl. 101 (2001) 45-53, arXiv:hep-th/0102012; P. Ramond,  {\it Algebraic Dreams }, Contribution to Francqui Foundation Meeting in the honor of Marc Henneaux, October 2001, Brussels, arXiv:hep-th/0112261;  L. Brink, P. Ramond and X. Xiong, {\it Supersymmetry and Euler Multiplets}, JHEP 0210 (2002) 058, arXiv:hep-th/0207253.

\end